\documentclass{aa}  

\usepackage{graphicx}
\usepackage{multirow}
\usepackage{color}
\usepackage{txfonts}
\newcommand{\speed}[1]{#1 km~s${}^{-1}$}

\newcommand{\nfig}[1]{Figure~\ref{#1}}
\newcommand{\nfigs}[1]{Figs.~\ref{#1}}
\newcommand{\fig}[1]{Fig.~\ref{#1}}
\newcommand{\ntab}[1]{Table~\ref{#1}}
\begin{document} 

   \title{Coronal Condensation as the Source of Transition Region Supersonic Downflows above a Sunspot}
\titlerunning{Transition Region Supersonic Downflows}

   \author{Hechao Chen 
          \inst{1}
          \and
          Hui Tian \inst{1,2}
          \and
          Leping Li \inst{2,3}
          \and
          Hardi Peter \inst{4}
          \and
          Lakshmi Pradeep Chitta \inst{4}
          \and
          Zhenyong Hou \inst{1}                              
          }

   \institute{School of Earth and Space Sciences, Peking University, Beijing, 100871, China. \\
         \email{huitian@pku.edu.cn}
         \and
             {Key Laboratory of Solar Activity, National Astronomical Observatories, Chinese Academy of Sciences, Beijing, 100101, China.}
              \email{lepingli@nao.cas.cn}
        \and
        {University of Chinese Academy of Sciences, Beijing, 100049, China} 
        \and
        {Max Planck Institute for Solar System Research, D-37077 Göttingen, Germany}                    
            }



  \abstract
  {Plasma loops {or plumes} rooted in sunspot umbrae often harbor downflows with speeds of ~\speed{100}. These downflows are supersonic at transition region temperatures of $\sim$ 0.1 MK. The source of these flows is not well understood.} 
   {We aim to investigate the source of sunspot supersonic downflows (SSDs) in active region (AR) 12740 using simultaneous spectroscopic and imaging observations.
   }
   {We identified SSD events from multiple raster scans of a sunspot by the \textit{Interface Region Imaging Spectrograph}, and calculated the electron densities, mass fluxes and velocities of these SSDs. The extreme-ultraviolet (EUV) images provided by the Atmospheric Imaging Assembly (AIA) onboard the \textit{Solar Dynamics Observatory} and the EUV Imager (EUVI) onboard the \textit{Solar Terrestrial Relations Observatory} (STEREO) were employed to investigate the origin of these SSDs and their associated coronal rain.}
{Almost all the identified SSDs appear at the footpoints of sunspot plumes and are temporally associated with {appearance} of {chromospheric bright dots inside the sunspot umbra}. 
{Dual-perspective EUV imaging observations reveal a large-scale closed magnetic loop system spanning the sunspot region and a remote region. We observed that the SSDs are caused by repeated coronal rain that forms and flows along these closed magnetic loops toward the sunspot. One episode of coronal rain clearly indicates that reconnection near a coronal X-shaped structure first leads to the formation of a magnetic dip.}  Subsequently, hot coronal plasma catastrophically cools from $\sim$$2$ MK in the dip region via thermal instability. This results in the formation of a transient prominence in the dip, from which the cool gas mostly slides into the sunspot along inclined magnetic fields under the gravity. This drainage process manifests as a continuous rain flow, which lasts for $\sim$2 hrs and concurrently results in a nearly steady SSD event.
The total mass of condensation ($ {1.3} \times 10^{14}$ g) and condensation rate  ($ {1.5} \times 10^{10}$ g s$^{-1}$) in the dip region were found to be sufficient enough to sustain this long-lived SSD event, which has a mass transport rate of  {$0.7-1.2\times 10^{10}$ g s$^{-1}$}. 
}
  { {Our results demonstrate that coronal condensation in magnetic dips can result in the quasi-steady sunspot supersonic downflows.} 
  }

   \keywords{Sun: sunspots --
                Sun: magnetic fields --
                Sun: chromosphere --
                Sun: transition region --
                Sun: corona
               }

   \maketitle
%

\section{Introduction}
High-resolution imaging and spectroscopic observations in the extreme ultraviolet (EUV) and far-ultraviolet (FUV) passbands have revealed many types of fine structures and dynamic events in the upper atmosphere above sunspots \citep[see a recent review by][]{2017RAA....17..110T}.  
One type of such phenomena are downflows harbored by plasma loops rooted in sunspots. These flows were first discovered as strongly redshifted secondary emission peaks at supersonic speeds in the transition region (TR) spectra in the 1980s \citep[e.g.,][]{1982SoPh...77...77D,1982SoPh...81..253N,1990Ap&SS.170..135B,1993ApJ...412..865G}. 

Early observations revealed that these sunspot supersonic downflows (SSDs) are quasi-steady, and that they tend to form at the footpoints of sunspot plumes \citep{1988ApJ...334.1066K,2001ApJ...552L..77B,2004ApJ...612.1193B}. 
With the launch of the \textit{Interface Region Imaging Spectrograph} \citep[IRIS,][]{2014SoPh..289.2733D}, SSDs in TR lines have recently received more and more attention.
A statistical investigation by \citet{2018ApJ...859..158S} revealed that: (1) SSDs commonly exist in the majority {($80 \%$)} of sunspots; (2) apart from redshifts in TR spectral lines, half of the SSD events have strong redshifted signatures also in chromospheric spectral lines. Thus, these SSDs likely play an important role in the {chromosphere-corona} mass cycle of the sunspot atmosphere. In TR spectral observations of IRIS, SSDs are found to be often associated with two types of spectra: (1) spectral lines that exhibit a broad enhancement at the red wings \citep[e.g.,][]{2014ApJ...789L..42K,2020SoPh..295...53I} or (2) spectral lines with a secondary emission peak that clearly separates from the near primary component \citep[e.g.,][]{2014ApJ...786..137T,2015A&A...582A.116S,2020A&A...640A.120N,2020A&A...636A..35N}. 
These above-mentioned studies show that the Doppler velocity, electron density, and mass flux of TR SSDs are generally in the range of \speed{$50-200$}, 10$^{9.48}-$10$^{10.86}$ cm$^{-3}$, and $10^{-7.5}-10^{-6}$ g cm$^{-2}$ s$^{-1}$, respectively.
These SSDs eventually can impart their energy into the lower atmosphere of sunspots as their speeds decrease from the supersonic to a sub-sonic state \citep{2015A&A...582A.116S,2016A&A...587A..20C,2021ApJ...916....5S}.

{Considering the time variation of their associated TR spectra, SSDs could be intermittent or long-lived. 
Intermittent SSDs, such as those observed by \citet{2014ApJ...789L..42K}, have been linked to impulsive coronal rain \citep{2020SoPh..295...53I} that may result from coronal condensation in the coronal part of the loops via thermal non-equilibrium {\citep[TNE, ][]{1991ApJ...378..372A,2019ApJ...884...68K}} and/or thermal instability  {\citep{1953ApJ...117..431P,1965ApJ...142..531F}} \footnote{{TNE and thermal instability are two important physical mechanisms to explain coronal rain and prominence formation in observations. For more details, refer to \citet{2019SoPh..294..173K} and \citet{2020PPCF...62a4016A}}.} The bursty (of the order 20 s) behavior of their spectra is naturally explained by the intermittent and clumpy character of coronal rain, as often seen in simulations \citep[e.g.,][]{2003A&A...411..605M,2004A&A...424..289M} and observations \citep[e.g.,][]{2015ApJ...806...81A}.
The origin of the long-lived and steady SSDs is less clear. In their spectra, the intensities and Doppler velocities of the downflow components often remain relatively stable on a timescale of at least $\sim$1 hour \citep{2014ApJ...786..137T,2015A&A...582A.116S}, meaning that a substantial and stable mass supply is required to sustain such SSDs.  \citet{2015A&A...582A.116S} and \citet{2016A&A...587A..20C} made a rough calculation and found that if all the material comes from a usual quiescent coronal loop undergoing catastrophic cooling near its apex, it would not have enough {plasma} to {sustain these steady downflows}. Hence, these authors speculated that these long-lived steady SSDs may be fed by siphon flows {from the other (conjugate) footpoint}.
}

{
The measured densities and velocities of long-lived steady SSDs {are similar to} those of coronal rain events \citep[e.g.,][]{2001SoPh..198..325S,2012ApJ...745..152A,2012SoPh..280..457A,2015ApJ...806...81A,2015A&A...582A.116S}. {Using} chromospheric observations at a sub-arcsecond spatial resolution, \citet{2015ApJ...806...81A} have found that in AR closed coronal loops, intermittent coronal rain clumps can be reshaped as more persistent and continuous rain flows at chromospheric heights, perhaps due to funnel{ing of magnetic fields} loops. These clues imply the possibility that both the intermittent and stable SSDs might have a common origin: flows of condensation forming above sunspots due to thermal instability and/or TNE. 
} 

{\citet{2018ApJ...864L...4L} have reported one category of quiescent coronal rain on the solar limb, where magnetic reconnection facilitates coronal condensation occurring in high-lying magnetic structures via the formation of magnetic dips.} Compared to the traditional condensation models \citep[e.g.,][]{2013ApJ...773...94M,2015ApJ...807..142F,2016ApJ...823...22X} that are based primarily on TNE and/or thermal instability in usual coronal loops, \citet{2018ApJ...864L...4L}'s coronal rain scenario highlights the important role of reconnection or reconfiguration of field lines in the onset of coronal condensation. With the presence of magnetic dips, plasma condensed in the hot corona via thermal instability and/or TNE can accumulate as (transient) solar prominences/filaments. {In this process,} cool {plasma} can drain to the solar surface along the legs of newly reconnected field lines, forming persistent and continuous coronal rain flows. {Although {the source of hotter material that initially feed the observed condensations remains unclear,} a series of similar observations revealed that continuous coronal rain flows {from such condensations} can persist over several to ten hours \citep[e.g.,][]{2019ApJ...884...34L,2020ApJ...905...26L}. If such long-lasting rain flows persistently fall into a sunspot along coronal loops, it might result in long-lived steady SSDs.}

{Recently, a He~{\sc{i}} spectropolarimetric study of SSDs from \citet{2021ApJ...916....5S} {further supports} this hypothesis. In their observations, after a coronal catastrophic cooling event occurring above a large sunspot, a cloud-like filament soon forms at the apex (or dips) of an active-region magnetic loop system and some portion of cool material subsequently drains towards the sunspot as blob- and stream-like rain flows, causing a chromospheric SSD event in simultaneous spectral observations \citep[see also][]{2016ApJ...833....5S}. However, their single (32-minute) slit scan at He~{\sc{i}} 10830 \AA \  triplet impeded them to address the time variation of SSD spectra at different {spectral lines}.
}

Here, we investigate a series of SSD events above NOAA active region (AR) 12740, using observations from IRIS, the New Vacuum Solar Telescope \citep[NVST;][]{2014RAA....14..705L}, the \textit{Solar Dynamics Observatory} (SDO), and \textit{Solar Terrestrial Relations Observatory} \citep[STEREO,][]{2004SPIE.5171..111W} on 2019 May 8--9. The joint observations obtained from multiple vantage points make them ideal ones for an investigation of the origin of SSDs. We find that these SSDs are closely related to repeated coronal ran activities, and provide strong evidence that reconnection-facilitated coronal condensation in high-lying magnetic dips can induce the long-lived steady SSD event. We describe the observational data in Section 2, present the main results in Section 3, and discuss our results in Sections 4. A brief conclusion is given in Section 5.
\par
\section{Observations and data}
\subsection{IRIS rasters and spectra}
IRIS has one near-ultraviolet (NUV; 2782.7–2835.1 \AA) and two far-ultraviolet channels (FUV; 1331.7–1358.4 \AA \ and 1389.0–1407.0 \AA), which can provide simultaneous images and spectra of the temperature minimum, chromosphere, TR, and corona \citep{2014SoPh..289.2733D}. During 2019 May 7-8, IRIS observed NOAA AR 12740 near the disk centre and performed five repeated raster scans. The field-of-view (FOV) of the slit-jaw images (SJI) for each raster is shown in \fig{fig1} (a). Note that the first four repeats were taken in the very large dense 320-step raster mode, while the last episode was taken in the 4-step raster mode (hereafter, we term them as Rasters 1-5, respectively). More information of these raster scans is listed in \ntab{tab1}. 
\par
To identify and characterize SSDs we employed the calibrated Level-2 IRIS spectral data, which has already been processed with dark current subtraction, flat fielding, orbital variation and geometrical corrections \citep{2014SoPh..289.2733D}. The following spectral lines were used in this study: (1) Si~{\sc{iv}} 1394/1403 \AA \ and O~{\sc{iv}} 1400/1401 \AA \ lines formed in the TR with a typical formation temperature of around 0.08 and 0.15 MK, respectively; (2) Mg~{\sc{ii}} h 2803 \AA \ and Mg~{\sc{ii}} k 2796 \AA \ lines with a formation temperature of 0.01 MK in the chromosphere; and (3) C~{\sc{ii}} 1334/1335 \AA \ lines with a formation temperature of around 0.25 MK, in the upper chromosphere or lower TR. Spectral lines from nearby neutral and singly ionized atoms, e.g., C~{\sc{i}}, O~{\sc{i}}, Ni~{\sc{ii}}, S~{\sc{i}}, Mn~{\sc{i}}, in a quiet-Sun region were used for the absolute wavelength calibration \citep[e.g.,][]{2016ApJ...824...96T,2016ApJ...829L..30H}.

\subsection{Imaging observations from the NVST, SDO, and STEREO-A}
To investigate the atmospheric response of SSDs at different heights, we also analyzed data from imaging observations with the NVST, SDO, IRIS, and STEREO-A. NVST is a 1-m ground-based vacuum solar telescope in China, which can observe fine structures in the lower atmosphere of the Sun \citep{2020ScChE..63.1656Y}. On 2019 May 8, NVST tracked the AR 12740 during the period of 01:10 $-$ 05:03 UT, and provided Level 1+ high-resolution images of the H$_{\alpha}$ 6562.8 \AA, H$_{\alpha}$ line wings at $\pm$ 0.4 \AA, and TiO. These images have been calibrated through dark current and flat field corrections, and were further reconstructed by a speckle masking method \citep{2016NewA...49....8X}. In addition, EUV images from the Atmospheric Imaging Assembly \citep[AIA;][]{2012SoPh..275...17L} onboard the SDO and 1400 \AA\, slit-jaw images  from the IRIS were utilized. All these images were co-aligned by finding common features observed in different channels. In addition, the 195 \AA, 304 \AA, and 171 \AA \ images from the Extreme Ultraviolet Imager \citep[EUVI;][]{2004SPIE.5171..111W} onboard the STEREO-A spacecraft were also analzyed to investigate the SSD-related coronal activity from a different perspective (see \fig{fig1} (b)). {The spatial resolution of all the images involved is given as follow, i.e. NVST H$_{\alpha}$ images $\sim$0.165$^{\prime\prime}$, NVST TiO images $\sim$0.052$^{\prime\prime}$, AIA EUV images $\sim$1.5$^{\prime\prime}$, IRIS slitjaw images $\sim$0.4$^{\prime\prime}$, and STEREO EUVI images $\sim$2.5$^{\prime\prime}$.}
  \begin{table*}[]
\caption{Detailed information of the IRIS Raster Scans and supersonic downflows.}
\label{tab1}
\resizebox{\textwidth}{!}{
\begin{tabular}{ccccccclcccccc}
\multicolumn{1}{l}{} & \multicolumn{1}{l}{} & \multicolumn{1}{l}{} & \multicolumn{1}{l}{} & \multicolumn{1}{l}{} & \multicolumn{1}{l}{} & \multicolumn{1}{l}{} &  & \multicolumn{1}{l}{} & \multicolumn{1}{l}{} & \multicolumn{1}{l}{} & \multicolumn{1}{l}{} & \multicolumn{1}{l}{} & \multicolumn{1}{l}{}\\ \hline
\multicolumn{7}{c}{Spectrograph Raster Scans over AR 12740} &  & \multicolumn{6}{c}{Supersonic downflows above the sunspot} \\ \cline{1-7} \cline{9-14} 
Raster Number & Date/Time & Center & FOV & Steps & Step Cadence & Exposure Time &  & \multicolumn{3}{c}{Region A} & \multicolumn{3}{c}{Region B} \\ \cline{9-11} \cline{11-14} 
(\#) & (UT) & (") & (") &   & (s) & (s) &  & 
\multicolumn{1}{c}{Detection?} &\multicolumn{1}{c}{Duration} &\multicolumn{1}{c}{log$(n_e)$} & \multicolumn{1}{c}{Detection?} &\multicolumn{1}{c}{Duration} &\multicolumn{1}{c}{log$(n_e)$} \\ \multirow{2}{*}{}&   &  &  &  &  &  &  &    & (s) &  (cm$^{-3}) $ &    & (s) & (cm$^{-3})$\\ \cline{1-7} \cline{9-14}\\
\multirow{2}{*}{1} & \multirow{2}{*}{07/23:00:50 - 08/01:53:40} & \multirow{2}{*}{-459, 213} & \multirow{2}{*}{112  $\times$ 175} & \multirow{2}{*}{320  $\times$ 0.35"} & \multirow{2}{*}{16.2} & \multirow{2}{*}{14.9} &  & \multirow{2}{*}{Yes} & \multirow{2}{*}{23:48:06-23:52:41} & \multirow{2}{*}{9.88 $\pm \ ^{0.42}_{0.95}$} & \multirow{2}{*}{No} & \multirow{2}{*}{-:-:-} & \multirow{2}{*}{$-$} \\
 &  &  &  &  &  &  &  &  &  &  &  \\
 \multirow{2}{*}{2} & \multirow{2}{*}{07/23:00:50 - 08/01:53:40} & \multirow{2}{*}{-459, 213} & \multirow{2}{*}{112  $\times$ 175} & \multirow{2}{*}{320  $\times$ 0.35"} & \multirow{2}{*}{16.2} & \multirow{2}{*}{14.9} &  & \multirow{2}{*}{Yes} & \multirow{2}{*}{01:08:02-01:19:06} & \multirow{2}{*}{10.33 $\pm \ ^{0.50}_{0.93}$} & \multirow{2}{*}{No} & \multirow{2}{*}{-:-:-} & \multirow{2}{*}{$-$} \\
&  &  &  &  &  &  &  &  &  &  &  \\
\multirow{2}{*}{3} & \multirow{2}{*}{08/02:01:58 - 08/03:28:23} & \multirow{2}{*}{-430, 214} & \multirow{2}{*}{112  $\times$ 175} & \multirow{2}{*}{320  $\times$ 0.35"} & \multirow{2}{*}{16.2} & \multirow{2}{*}{14.9} &  & \multirow{2}{*}{Yes} & \multirow{2}{*}{02:43:17-02:50:02} & \multirow{2}{*}{10.47 $\pm \ ^{0.39}_{0.69}$} & \multirow{2}{*}{Yes} & \multirow{2}{*}{02:51:56-02:54:38} & \multirow{2}{*}{$-$} \\
&  &  &  &  &  &  &  &  &  &  &  \\
\multirow{2}{*}{4} & \multirow{2}{*}{08/05:09:50 - 08/06:36:15} & \multirow{2}{*}{-421, 196} & \multirow{2}{*}{112  $\times$ 175} & \multirow{2}{*}{320  $\times$ 0.35"} & \multirow{2}{*}{16.2} & \multirow{2}{*}{14.9} &  & \multirow{2}{*}{No} & \multirow{2}{*}{-:-:-} & \multirow{2}{*}{$-$} & \multirow{2}{*}{Yes} & \multirow{2}{*}{06:12:45-06:18:42} & \multirow{2}{*}{10.52 $\pm \ ^{0.48}_{0.93}$}  \\
 &  &  &  &  &  &  &  &  &  &  &  \\
\multirow{2}{*}{5} & \multirow{2}{*}{08/16:29:36 - 08/19:30:50} & \multirow{2}{*}{-306, 188} & \multirow{2}{*}{3  $\times$ 119} & \multirow{2}{*}{4  $\times$ 1"} & \multirow{2}{*}{8.8} & \multirow{2}{*}{7.9} &  & \multirow{2}{*}{Yes} & \multirow{2}{*}{18:55:12:-19:30:33} & \multirow{2}{*}{10.44 $\pm \ ^{0.41}_{0.64}$} & \multirow{2}{*}{No obs} & \multirow{2}{*}{-:-:-} & \multirow{2}{*}{$-$} \\
 &  &  &  &  &  &  &  &  &  &  &  \\ \hline
\multicolumn{1}{l}{} & \multicolumn{1}{l}{} & \multicolumn{1}{l}{} & \multicolumn{1}{l}{} & \multicolumn{1}{l}{} & \multicolumn{1}{l}{} & \multicolumn{1}{l}{} &  & \multicolumn{1}{l}{} & \multicolumn{1}{l}{} & \multicolumn{1}{l}{} & \multicolumn{1}{l}{} & \multicolumn{1}{l}{} & \multicolumn{1}{l}{}
\end{tabular}
}
\tablefoot{The observation dates are 2019 May 7--8. In columns 9 and 12, the durations are the time periods when the SSDs were detected in regions A and B during the raster scans, respectively. The symbol ``-:-:-" means that no SSDs were detected as the slit scanned this region. {The symbol ``$-$" means that no SSDs were detected as the slit scanned this region or the O~{\sc{iv}} 1400/1401 \AA \ line pair are too weak for density diagnostics.}}
\end{table*}

\section{Analysis and results}
\subsection{Sunspot supersonic downflows detected by IRIS}
\subsubsection{Spatial extent and temporal evolution}
\begin{figure*}
   \centering
   \includegraphics[width=.75\hsize]{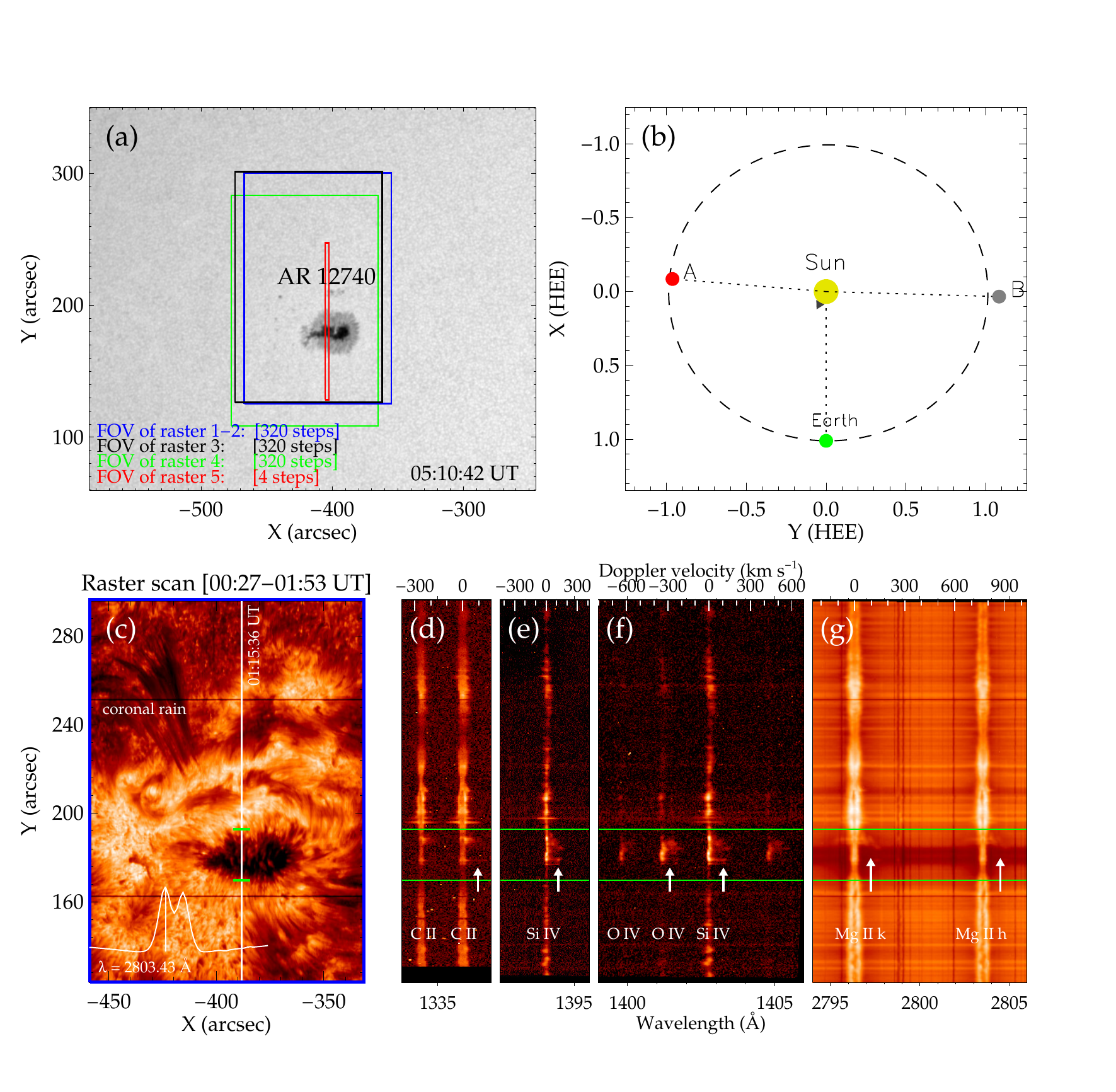}
      \caption{Overview of the region-of-interest. (a) HMI continuum image of NOAA AR 12740. The rectangles outlined by the blue, black, green, and red lines indicate the field-of-views (FOVs) of spectral observations for IRIS rasters 1-2, 3, 4 and 5, respectively. (b) Positions of the STEREO-A spacecraft in the ecliptic plane at 05:30 UT on 2019 May 8. The target region was simultaneously observed by the STEREO-A and SDO from two perspectives. (c) An image of IRIS raster scan at Mg~{\sc{ii}} k 2803.43 \AA. The vertical white line indicates the IRIS slit location. (d)-(g) IRIS spectra along the slit shown in (c). The green horizontal lines mark the location of the sunspot. The two white arrows in (g) indicate the downflow components of the Mg~{\sc{ii}} lines. An animation of this figure is available online.
              }
         \label{fig1}
   \end{figure*}

As mentioned above, from 2019 May 7 to 8, through five raster scans IRIS detected clear spectral signatures of SSDs as the slit scanned the sunspot region. As presented in \fig{fig1} (c) and its associated animation, the Mg~{\sc{ii}} k$_{2}$ and h$_{2}$ images reconstructed from the 320-step raster scans of IRIS in \fig{fig1} (c) reveal that elongated coronal rain flows falling into the sunspot. One example of SSD events is shown in \nfigs{fig1} (d-g). We can see that the SSDs manifest themselves as obvious emission enhancements in the red wings of C~{\sc{ii}}, Si~{\sc{iv}}, O~{\sc{iv}}, and Mg~{\sc{ii}} lines. In \fig{fig2} (a), an overview of the sunspot region at different atmospheric layers is presented. Note that the sunspot can be divided into two parts by a developing light bridge. For brevity, hereafter, we term the eastern (western) part of the umbra as region A (B). 
\par
Through scrutinizing IRIS spectra, we found clear signatures of SSD events occurring in both regions A and B. At a temperature of 0.08 MK, i.e. in the source region of Si~{\sc{iv}}, the sound speed is of the order of \speed{50}. This underlines that the velocities found in the SSDs are very high. In \ntab{tab1}, we list the time periods during which SSDs were detected. To examine the spatial extent and temporal evolution of SSDs, for each raster we summed the red wing of the Si~{\sc{iv}} 1403 \AA \ line profile over the spectral positions with Doppler shifts {larger than 50 km s$^{-1}$ and smaller than \speed{300}}, and obtained a map of the downflow intensity. \nfig{fig2} (b) presents such maps for Rasters 1-4. 
Regarding the spatial structures of SSDs, a strand-like morphology is found in both regions A and B \citep[see similar features also in][]{2020A&A...640A.120N}. The projected length (width) of the downflow structures is about 6$^{\prime\prime}-$19$^{\prime\prime}$ (2$^{\prime\prime}-$3$^{\prime\prime}$), comparable to the upper limit of the length (width) of coronal rain clumps \citep{2015ApJ...806...81A}. 
A comparison of these structures with the AIA 171 \AA\ image in \nfig{fig2} (a) suggests that these downflows are magnetically confined at the footpoints of sunspot plumes \citep[see also][]{2016A&A...587A..20C,2020A&A...640A.120N,2020A&A...636A..35N}.
Regarding their time variation, we can see that the stand-like downflow structures in region A  {are} enhanced from Raster 1 to 3,  disappear in Raster 4, and then recur during Raster 5 (see also \fig{fig4}). A similar structure in region B first appear during Raster 3, and undergo an enhancement during Raster 4. 
{These results reveal that SSDs exhibit a relatively active temporal evolution \citep[see also][]{2020A&A...640A.120N}, which are quite similar to the observed behavior of usual quiescent coronal rain flows occurring in AR closed loops \citep[e.g.,][]{2015ApJ...806...81A}.}  
\subsubsection{Spectral line parameters and mass flux}
   \begin{figure*}
   \centering
   \includegraphics[width=0.8\hsize]{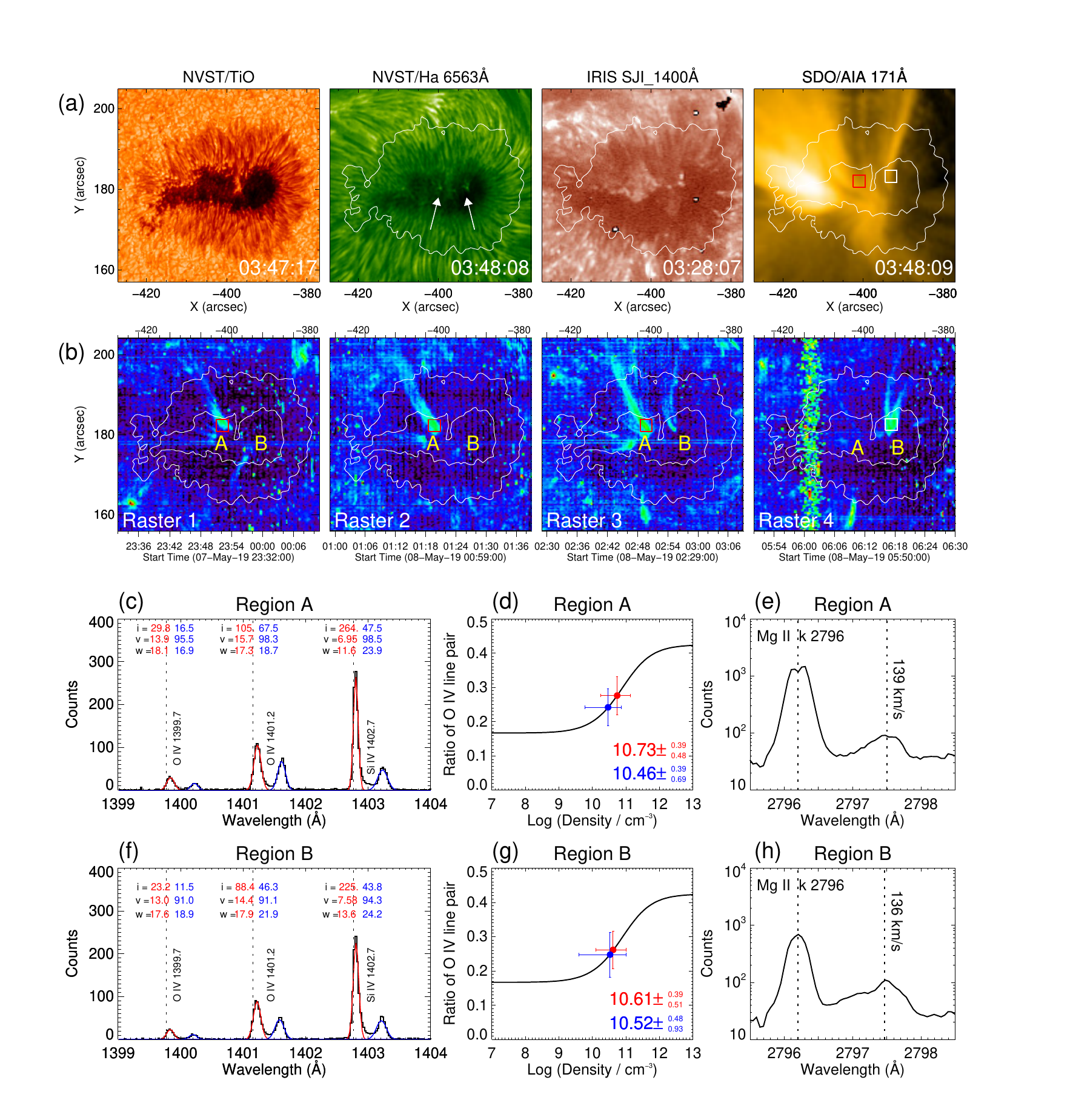}
      \caption{Multi-wavelength diagnostics of sunspot supersonic downflows. (a) NVST TiO and 6563 \AA,  IRIS SJI 1400 \AA, and  AIA 171 \AA \ images of the target sunspot. The white contours outline the outer boundaries of the sunspot umbra and penumbra. (b) Spatial extent of SSDs as shown in maps of the downflow intensity {integrated in the velocity range of 50 km s$^{-1}$ to \speed{300} }. (c) Average spectrum in the small red box (3$^{\prime\prime}$$\times$3$^{\prime\prime}$) shown in panel (b) (Raster 3). A six-component Gaussian fit to the spectrum is shown as the red (primary component) and blue lines (secondary component). The fitting parameters, including the peak intensity (i), Doppler shift (v), and line width (w), are also given for each line. (d) The black curve shows the theoretical relationship between the electron density and the ratio of the O~{\sc{iv}} doublet. The blue and red solid circles (error bars) represent the density diagnostic results (associated uncertainties) of the secondary and primary components in region A, respectively. (e) Average Mg~{\sc{ii}} k 2796 \AA\, line profile in region A (Raster 3). The two vertical dotted lines mark the two components. (f)-(h): Same as (c)-(e) but for the SSDs in region B. The spectrum in (f) was averaged over the small white box (3$^{\prime\prime}$$\times$3$^{\prime\prime}$) shown in panel (b) (Raster 4).}
         \label{fig2}
   \end{figure*}
We derive spectral line parameters of SSD events focusing on a small spectral window from 1399 \AA \ to 1404 \AA. This spectral window contains the Si~{\sc{iv}} 1403 \AA \ line, and the O~{\sc{iv}} 1400/1401 \AA \ line pair whose ratio is sensitive to the electron density of the TR plasma. Examples of spectra in the selected window are displayed in \nfigs{fig2} (c) and (f). These spectra have been averaged over the red box in region A during Raster 3 and the white box in region B during Raster 4, respectively. Due to the existence of SSDs, {a downflow component appears as a secondary emission peak} in the red wing of each of the O~{\sc{iv}} 1400 \AA, O~{\sc{iv}} 1401 \AA, and Si~{\sc{iv}} 1403 \AA \ lines. As shown in \nfigs{fig2} (c) and (f), a six-component Gaussian function was applied to fit the spectra \citep[also see][]{2018ApJ...859..158S}. The fitting results of the primary and secondary components are plotted separately as the red and blue lines, respectively. The Doppler shifts, line widths and peak intensities of both the primary and secondary (downflow) components during different raster scans were then obtained. For the faster SSD component, the Doppler shifts of the O~{\sc{iv}} 1400/1401 \AA \ and Si~{\sc{iv}} 1403 \AA \ lines in the two regions appear to be very similar, with an average value of $\sim$\speed{95}. For the primary components, the average Doppler shifts of the O~{\sc{iv}} 1400/1401 \AA \ and Si~{\sc{iv}} 1403 \AA \ lines in the two regions are around \speed{14} and \speed{7}, respectively. {These redshifts are similar in magnitude to the well-known average redshifts of TR lines \citep[e.g.,][]{1993ApJ...408..735B,1997SoPh..175..349B,1998ApJS..114..151C,1999ApJ...522.1148P,1999ApJ...516..490P}.} It is worth mentioning that these SSDs were also detected in the Mg~{\sc{ii}} k 2796 \AA \  line. As shown in \nfigs{fig2} (e) and (h), an isolated downflow component also appears in the red wing of the Mg~{\sc{ii}} k line. The peak of this component corresponds to a red shift of $\sim$\speed{135}.
This corresponds to a flow significantly faster than seen in Si~{\sc{iv}} and O~{\sc{iv}} forming at higher temperatures. We will revisit this point in Section 4.1.
\par
The intensity ratio of the O~{\sc{iv}} 1400/1401 \AA \ line pair is known to be sensitive to electron density in the TR. Based on the CHIANTI atomic database 7.1 \citep{1997A&AS..125..149D,2013ApJ...763...86L}, the theoretical dependence of the intensity ratio of the O~{\sc{iv}} line pair on electron density was calculated. We estimated the electron densities, log$(n_\mathrm{e})$, of both the primary and downflow components {during each raster} from the measured intensity ratios of the O~{\sc{iv}} line pair{, and list them in \ntab{tab1}}. {Examples of electron density diagnostics in Regions A and B are shown in \nfigs{fig2} (d) and (g), respectively.} This will represent the density at temperatures of around 0.15 MK, i.e. in the source region of O~{\sc{iv}}, which is forming at temperatures almost a factor of two higher than Si~{\sc{iv}}. {For region A, the logarithmic density of downflow component varies in the range of 9.88$-$10.47 cm$^{-3}$ during Rasters 1$-$5 (see \ntab{tab1});} and the {average values} of the primary and downflow components are {10.69 $\pm \ ^{0.45}_{0.65}$ cm$^{-3}$ and  10.28 $\pm  \ ^{0.44}_{0.83}$ cm$^{-3}$, respectively. For region B, the logarithmic electron density of the primary and downflow components are only obtained in Raster 4, which are 10.61 $\pm  \ ^{0.39}_{0.51}$ cm$^{-3}$ and 10.52 $\pm \ ^{0.48}_{0.93}$ cm$^{-3}$, respectively.}
These values roughly agree with the estimated values found in the statistical work of \citet{2018ApJ...859..158S} {, and the density values of the downflow components are comparable to the typical electron density of coronal rain as well \citep[e.g.,][]{2015ApJ...806...81A}}.  

Assuming a He/H abundance ratio of 1:10 and complete ionization of the plasma, the mass flux ($\Delta m$) of SSDs can be computed as $1.2n_\mathrm{e} m_\mathrm{p}  v_\mathrm{SSD}$, where $m_\mathrm{p}$ denotes the mass of proton. We used the Doppler velocity of SSDs, $\sim$\speed{100}, as $v_\mathrm{SSD}$. As a result, the typical mass flux of SSDs was estimated to be $ {3.8} \times 10^{-7}$ g cm$^{-2}$ s$^{-1}$ in region A and $ {6.8} \times 10^{-7}$ g cm$^{-2}$ s$^{-1}$ in region B.
These values are comparable to the mass flux of a quasi-steady SSD event studied by \citet{2015A&A...582A.116S}.   

\subsection{Linking sunspot supersonic downflows to coronal rain flows}
\begin{figure*}
   \centering
   \includegraphics[width=0.7\hsize]{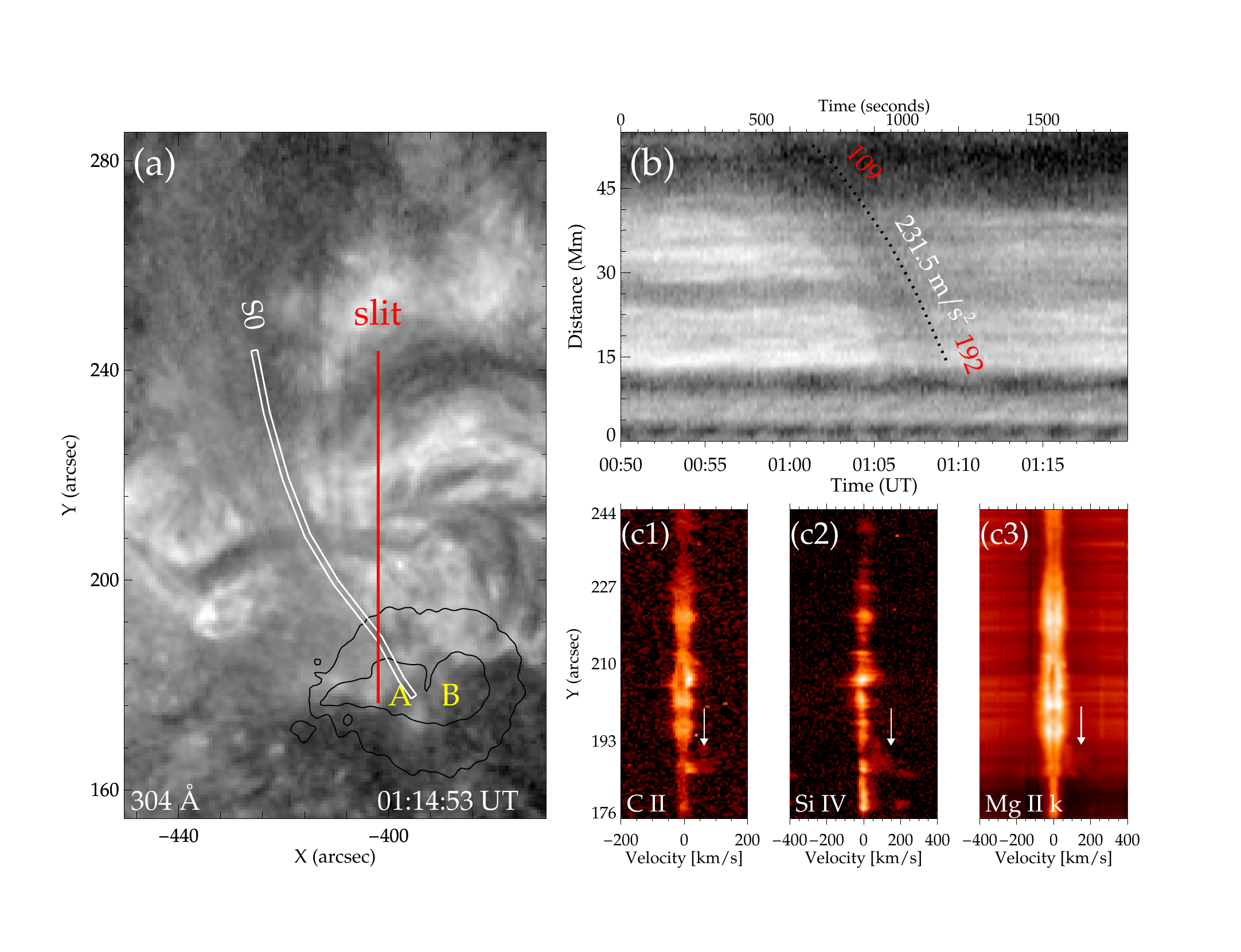}
      \caption{Temporal relation between coronal rain and sunspot supersonic downflows. (a) An AIA 304 \AA \ image showing the coronal rain falling into region A. The black contours outline the boundaries of the umbra and penumbra. (b) Space-time diagram of AIA 304 \AA \ intensity along the trajectory S0 shown in panel (a). The dark dotted line represents a parabolic fit to the track of one coronal rain flow occurring from 00:50 to 01:20 UT. (c) Spectral detector images of the C~{\sc{ii}} 1335 \AA \, Si~{\sc{iv}} 1403 \AA \, and Mg~{\sc{ii}} k 2796 \AA \ lines along the slit (red vertical line) shown in panel (a).  The white arrows mark the SSDs caused by the falling coronal rain along trajectory S0. An animation of this figure is available online.
              }
         \label{fig3}
   \end{figure*}

The simultaneous SDO/AIA and STEREO-A/EUVI observations reveal that repeated coronal rain events occur near the sunspot during the IRIS raster scans. 
To verify the close link between the SSDs and these coronal rain flows, we give a detailed investigation on their temporal and spatial correspondence.
\nfig{fig3} shows one episode of on-disk coronal rain between 00:50 UT to 01:20 UT. At the same time, IRIS (Raster 2) scanned the same region. The coronal rain events are best captured in the AIA 304 \AA \ image sequence, where we can see obvious cool materials draining downward to the sunspot (region A) along a set of trajectories in a funnel-like structure. In \fig{fig3} (b), we present the space-time diagram for one falling path (trajectory S0 in panel (a)). The parabolic track of the coronal rain flows indicates an acceleration of 230 m s$^{-2}$ ($\sim$5/6 of the solar gravity). The apparent velocity increases to $\sim$\speed{192} near the sunspot region. With simultaneous IRIS observations, these falling cool materials can be identified as a redshifted feature in the C~{\sc{ii}} 1334/1335 \AA, Si~{\sc{iv}} 1403 \AA, and Mg~{\sc{ii}} k 2796 \AA \ lines (see \nfigs{fig3} (c1-c3) and the associated animation). The coronal-rain-induced redshifted feature rapidly drifted southward along the slit with an increasing redshift (up to \speed{200}). Note that the instantaneous spectral redshift is comparable to the apparent falling velocity measured from the space-time diagram, suggesting that the falling path is aligned roughly 45 deg with respect to the line of sight. The largest redshift is much higher than the average redshift (\speed{$\sim$ 95}) of SSDs, likely because the latter was inferred from four spectra that separately averaged over 3$^{\prime\prime}$$\times$3$^{\prime\prime}$ small boxes in \fig{fig2} (b) during Rasters 1-4. 
\par
\begin{figure*}
   \centering
   \includegraphics[width=0.8\hsize]{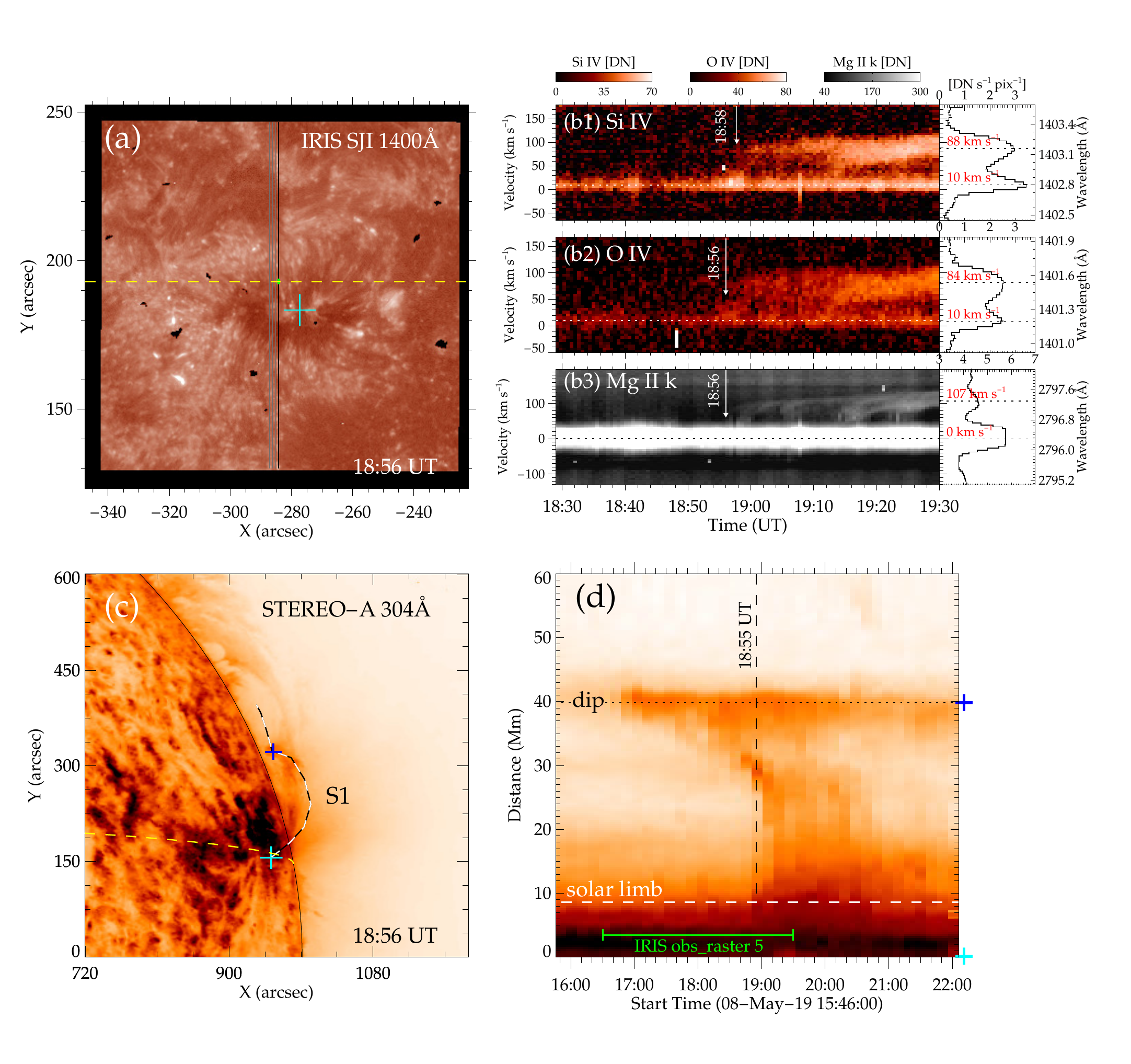}
      \caption{Sunspot supersonic downflows (SSDs) detected during IRIS Raster 5 and their associated coronal rain events. 
      (a) An IRIS SJI 1400 \AA \ image. The vertical lines indicate the four slit positions of the 4-step raster scans. SSDs were best detected at the last slit position (the black line). {(b1)-(b3)} Temporal evolution of the Si~{\sc{iv}} 1403 \AA, O~{\sc{iv}} 1401 \AA \,  {and Mg~{\sc{ii}} 2796 \AA \  line} profiles averaged within the section marked by the green line shown in (a). The time-averaged spectra are also plotted in the right, and the Doppler shifts of the primary and downflow components are given. The horizontal dotted lines indicate the wavelengths corresponding to the peak intensities of the two components. (c) The corresponding coronal rain event in an EUVI 304 \AA \ image from a different perspective. The (magnetic) dip region is marked by the blue plus sign. (d) Space-time diagram of EUVI 304 \AA \ intensity for the trajectory ``S1" shown in panel (c). The period of the IRIS 4-step rasters is marked by the green solid line at the bottom. In (a) and (c), the cyan plus signs mark the same heliocentric position after considering the solar rotation, and the yellow dashed line is the north 11.9 deg altitude line.}
         \label{fig4}
   \end{figure*}

\nfig{fig4} presents images taken during another episode of IRIS observations from 16:30 UT to 19:30 UT, i.e., Raster 5, together with simultaneous STEREO-A/EUVI observations. 
Since STEREO-A observed the Sun from a different perspective, the coronal rain events were recorded at the limb in EUVI images. During this period IRIS performed 4-step repeated raster scans above the sunspot, in region A. Obvious spectral signatures of the SSDs were detected at the 4th slit position shown in \fig{fig4} (a). We averaged the spectra of Si~{\sc{iv}} 1403 \AA,  O~{\sc{iv}} \AA, and {Mg~{\sc{ii}} k} within the section marked by the green line shown in \fig{fig4} (a), and plotted their temporal evolution in \nfigs{fig4} (b1)-(b3). Interestingly, the SSDs we detected here appear as a quasi-steady secondary redshifted emission peak in both TR and chromospheric lines, which remains relatively stable between 18:55 UT to 19:30 UT with a redshift of \speed{84$-$88} in TR lines and of \speed{107} in Mg~{\sc{ii}} k.
In the meantime, EUVI captured an episode of coronal rain draining into the sunspot region from a (magnetic) dip in the 304 \AA \ channel (\fig{fig4} {(c)} and the associated animation of \fig{fig5}). To better illustrate this process, a space-time diagram was constructed for the trajectory ``S1" shown in \fig{fig4} {(c)} , which passes through the dip region (marked by the green ``+" sign) and the sunspot region (marked by the cyan ``+" sign). Note that a cyan ``+" is also marked at the same heliocentric position in \fig{fig4} (a), which is located at the centre of the sunspot. From the space-time diagram shown in \fig{fig4} (d), we can see that the cool materials first appear in the dip region around 17:00 UT, and then drain into the sunspot during the period of 18:55 UT$-$ 21:00 UT (around 2 hrs). This is well in line with the period of SSD appearance in the TR spectral lines, as shown in \nfigs{fig4} {(b1)-(b3)} . This temporal correspondence further confirms that the SSD events detected in our observations originate from the mass drainage in coronal rain events. Unfortunately, this IRIS observation ended at 19:30 UT, whereas the mass drainage of coronal rain lasts for another 1.5 hrs.
Note that in \fig{fig4} {(b3), there seems to be an acceleration trend of the rain in time with rain produced later falling faster. This trend may also be present in the TR lines (see \nfigs{fig4} {(b1)-(b2)}), but due to their lower spatial and spectral resolution it is not as clear as in Mg~{\sc{ii}} k.}
\par
   \begin{figure*}
   \centering
   \includegraphics[width=0.65\hsize]{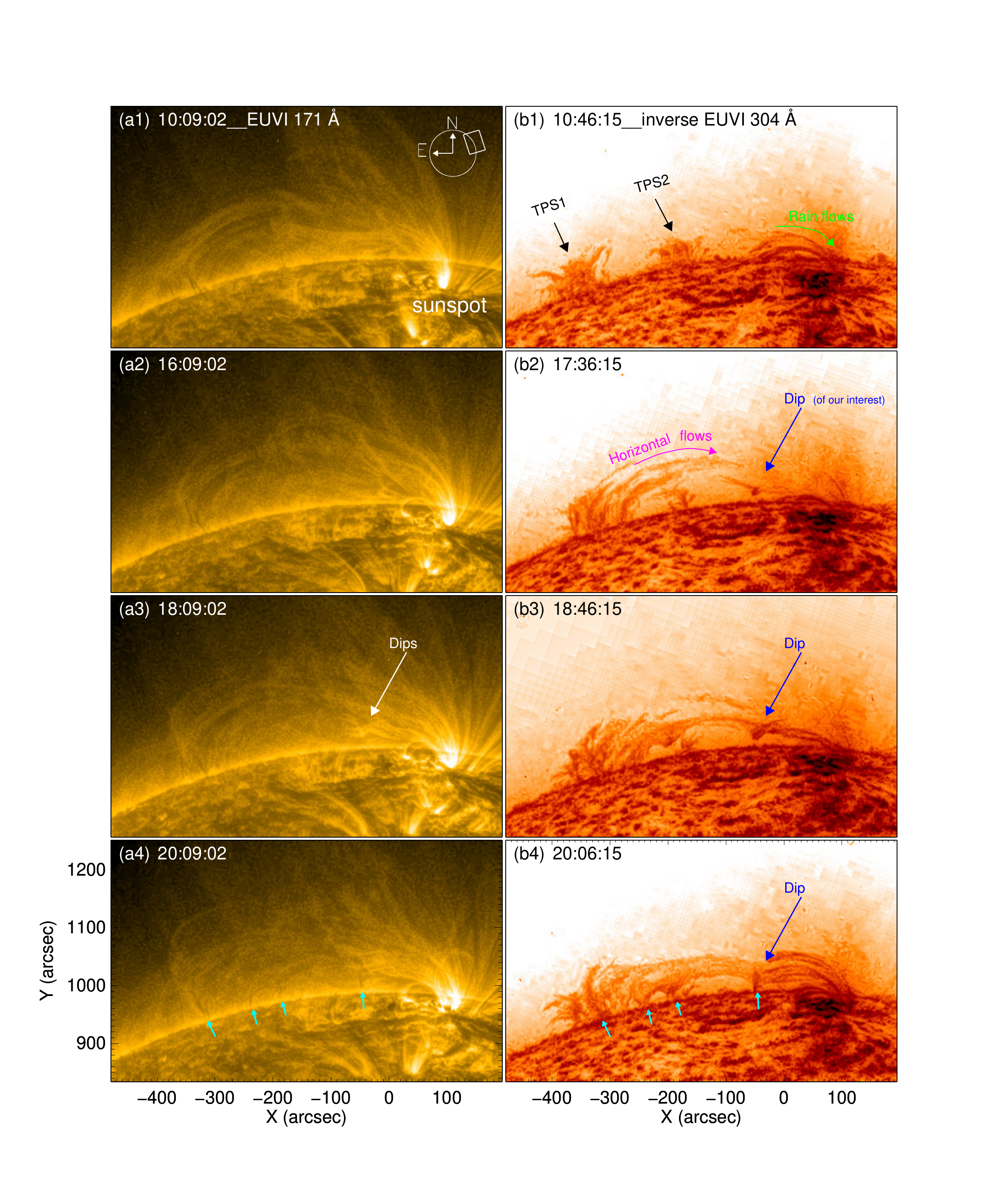}
      \caption{{Global view of coronal rain activity and forming prominence structures near the sunspot observed by STEREO-A/EUVI. (a1)-(a4) 171 \AA \ and (b1)-(b4) 304 \AA \ images are rotated counterclockwise by an angle of 70 deg, and the latter has an inverse color table. The solar disk and the FOV of these images are shown as the white circle and the rectangle, respectively. The black arrows denote two forming tower prominence structures (TPSs); the green curved arrow shows the trajectory of rain flows near the sunspot; the purple curved arrow denote the trajectory of a large-scale horizontal flow; the blue arrows mark the dip of our interest; the cyan arrows denote the feet structures of the formed quiescent prominence. Note that the EUVI images were enhanced by the MGN technique \citep{2014SoPh..289.2945M}.}}
         \label{figx}
   \end{figure*}
\subsection{Coronal rain activities and magnetic dips}
We then investigated the formation process of the coronal rain. 
We examined the AIA and EUVI observations, and confirmed the absence of flares near the sunspot. 
{\nfig{figx} shows a global picture of the coronal rain activity from the vantage point of STEREO-A. As shown in EUVI 171 and 304 \AA \ images, a large-scale closed magnetic loop system spans from remote regions to AR 12740, in which a quiescent prominence is gradually forming during 10:00$-$21:00 UT. In this course, a series of coronal rain also repeatedly form in the magnetic loop system and flow towards the sunspot region during the IRIS raster scans (Rasters 1$-$5). Considering their forming regions, these coronal rain flows can be characterized as three types. One type is typical quiescent coronal rain flows appearing near the west leg of the loop system, which repeatedly form in sunspot plumes and directly flow into the sunspot during the period of 01:00 $-$16:00 UT. The second type is episodes of large-scale {apparent} horizontal mass flows, {seen as spill-out events} from the nearby funnel or tower prominence structures \citep[e.g.,][]{2014IAUS..300..441L,2021NatAs...5...54A,2021A&A...647A.112Z} {that} flow towards the sunspot region along overlying large-scale coronal loops (as marked in \nfig{figx} (b2)).
A third type is continuous rain flows originating from a newly-formed magnetic dip region (as marked in \nfig{figx} (b)) during Raster 5. Note that a portion of large-scale horizontal mass flows seem to continuously inject cool material into this newly-formed dip region. }

\begin{figure*}
   \centering
  \includegraphics[width=0.6\hsize]{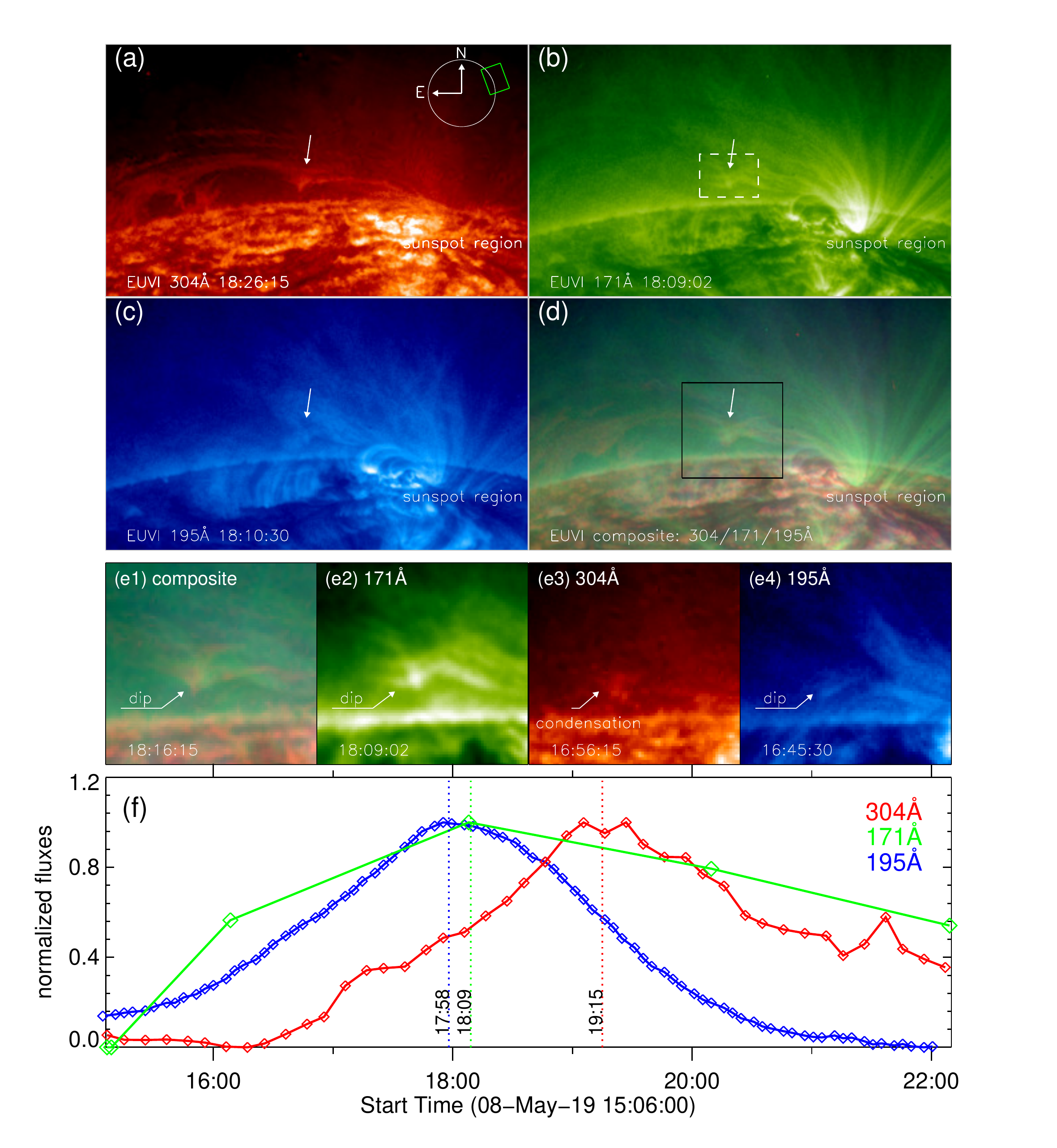}
      \caption{Coronal condensation observed by STEREO-A/EUVI during IRIS Raster 5. (a) 304 \AA, (b) 171 \AA, (c) 195 \AA, and (d) their composite images. These images are rotated counterclockwise by an angle of 70 deg. The solar disk and the FOV of these images are shown as the white circle and the green rectangle, respectively. The black rectangle in (d) denotes the FOV of (e).  (e) four selected close-in snapshots showing the evolution of the (magnetic) dip region at the onset of coronal condensation. White arrows mark the dip and associated condensation. (f) Temporal evolution of the 195 \AA \ (blue), 171  \AA \ (green), and 304  \AA \ (red) intensities integrated over the white box shown in panel (b).  The vertical blue, green, and red dotted lines mark the peak times of the three light curves, respectively. An animation of this figure is available online.
                   }
         \label{fig5}
   \end{figure*}

{Magnetic dips are region of upward concavity appearing upon field lines, in which the cold condensation can settle in and grow in size. Thus, the appearance of magnetic dips is thought to be important for the formation and maintenance of prominences/filaments \citep[e.g.,][]{2012ApJ...745L..21L,2015ASSL..415.....V,2018SoPh..293...93C}. }
From \fig{fig5} and its associated online animation, we can that a clear coronal condensation process occurs at the dip region of our interest. This process can be identified from the EUVI 304 \AA, 195 \AA, and 171 \AA \ images. The low-cadence (2 hr) EUVI 171 \AA \ images appear to suggest that the dip forms during the period of 16:06 UT$-$18:09 UT. Thanks to the higher-cadence EUVI 195 \AA \ (5 min) and 304 \AA \ (10 min) observations, we confirm that the dip first appears at 16:45:30 UT in the EUVI 195 \AA \ passband (see \fig{fig5} (e4)), and evident condensation starts subsequently at 16:56:15 UT (see \fig{fig5} (e3)). {This timing suggests the formation of the dip as a possible precursor of the onset of coronal condensation in the nearby region.} 
As cool materials gradually fill the dip region, continuous rain flows drain towards the sunspot along high-lying magnetic structures at about 18:46:15 UT. In the meantime, the dip region experiences a downward motion, {temporarily manifesting as one of the tower or pillar prominence feet (as marked by cyan arrows in \nfigs{figx} (a4) and (b4))}, which supports that some vertical feet of prominences may consist of piling-up magnetic dips \citep[e.g.,][]{2012ApJ...745L..21L,2015ApJ...814L..17S}. Subsequently, the cool materials also drain along a set of newly-formed magnetic arches linking the site below the dip region to the sunspot region around 20:46:15 UT, implying the generation of new loops through reconnection between the dip region and the lower-lying magnetic field. This is similar to the coronal condensation phenomena reported by \citet{2018ApJ...864L...4L} {and \citet{2019ApJ...887..137S}, in which reconnection seems to take part in.}
\nfig{fig5} (f) shows that the EUVI 195 \AA \ ($\sim$1.6 MK), 171 \AA \ ($\sim$0.8 MK), and 304 \AA \ ($\sim$0.08 MK) light curves in the dip region peak at 17:58 UT, 18:09 UT, and 19:15 UT, respectively. This indicates that coronal plasma in the dip region is cooling, as often observed in the process of prominence formation \citep[e.g.,][]{2012ApJ...758L..37B,2012ApJ...745L..21L} {and coronal rain formation \citep[e.g.,][]{2019ApJ...874L..33M,2019ApJ...884...34L,2019A&A...630A.123K,2020PPCF...62a4016A}. } 

   \begin{figure*}
   \centering
   \includegraphics[width=0.75\hsize]{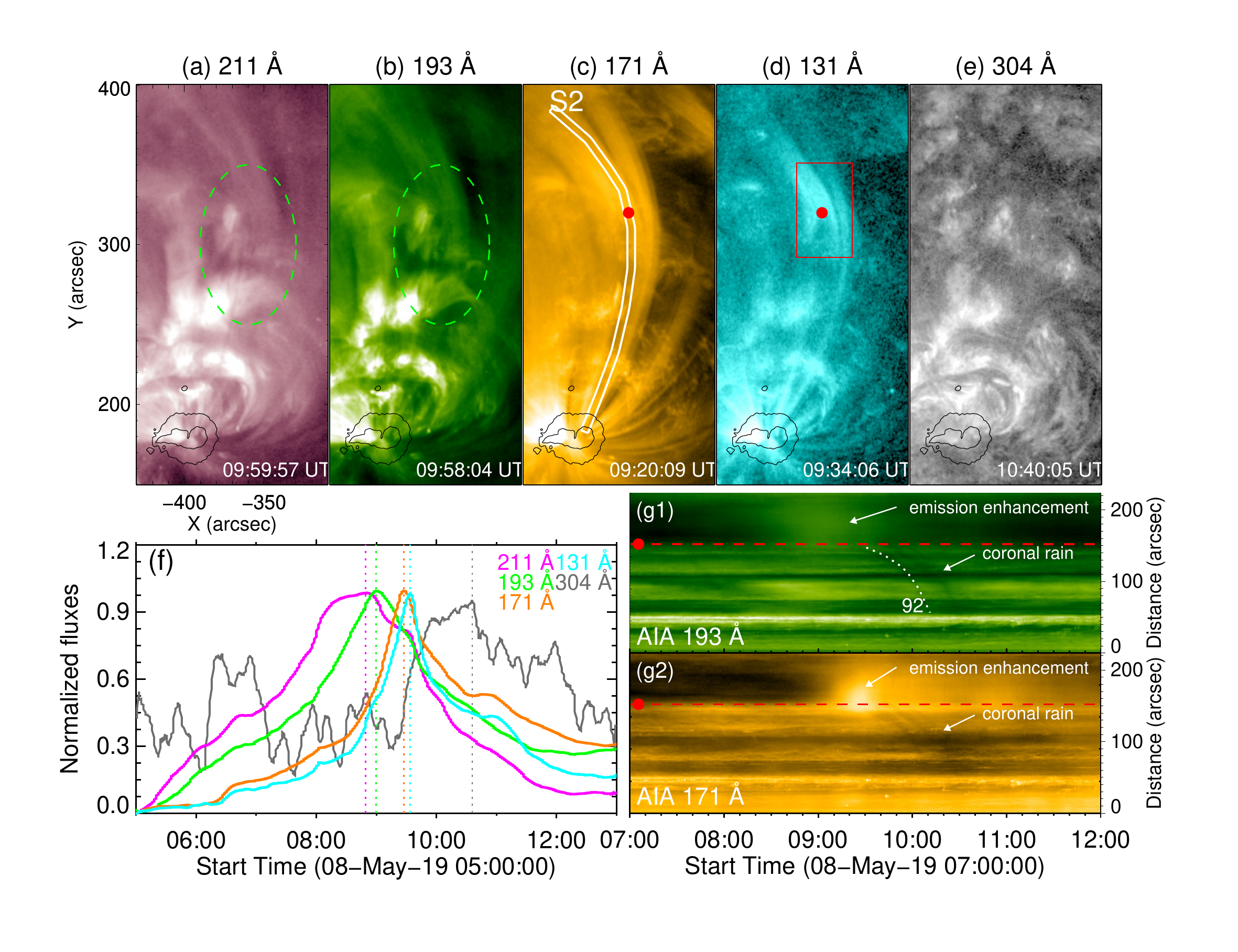}
      \caption{On-disk coronal condensation observed by AIA. (a) 211 \AA,
      (b) 193 \AA, (c) 171 \AA, (d) 131 \AA, and (e) 304 \AA \ images. The green dashed ellipses in (a) and (b) mark the location of the condensation; the red dots in (c) and (d) denote the location of a magetic dip region. (f) Normalized light curves of the AIA 211 \AA \ (purple), 193 \AA \ (green), 171 \AA \ (brown), 131 \AA \ (cyan), and 304 \AA \ (grey) channels in the red box shown in panel (d). (g) Space-time diagrams of AIA 193 \AA \ and 171 \AA  \  intensities for the trajectory ``S2" shown in panel (c). The white dotted line in (g1) represents a parabolic fit to the track of one coronal rain flow, and the red dash lines indicate the dip region.
      An animation of this figure is available online.
              }
         \label{fig6}
   \end{figure*}
\par 
The AIA observations show that similar condensation processes repeatedly take place near regions A and B, resulting in episodes of on-disk quiescent coronal rain events \citep[e.g.,][]{2012SoPh..280..457A,2021ApJ...910...82L}.  
\nfig{fig6} and its associated online animation present AIA observations for one episode of such coronal condensation. In 131 \AA \ ($\sim$0.6 MK) and 171 \AA \ ($\sim$0.8 MK) images, a localized emission enhancement  {is} first detected around 09:35 UT at the location of (-364$^{\prime \prime}$, 320$^{\prime \prime}$). This location (red dots in \fig{fig6} (c) and (d)) roughly coincides with a magnetic dip region in our PFSS extrapolation that will be presented in Section 4.2 (see \fig{fig10} (b)).
In 211 \AA \ ($\sim$2.0 MK) and 193 \AA \ ($\sim$1.6 MK) images, coronal rain clumps resulting from the coronal condensation subsequently appear around 09:50 UT (within the green eclipse), and then rapidly drain into region B along the loops. In comparison, the 304 \AA \ passband  {captures} more coronal rain flows during the same period due to its lower formation temperature ($\sim$0.08 MK). To better illustrate this on-disk condensation process, two space-time diagrams were constructed for the trajectory ``S2" shown in \fig{fig6} (c). The space-time diagrams shown in \fig{fig6} (g) reveal that cool materials appear after a localized enhancement in the AIA 193 \AA \ and 171 \AA \ images. This enhanced emission is likely caused by a local plasma density increase in the (magnetic) dip region.
To examine whether coronal condensation occurs, we plotted the temporal evolution of the AIA 211 \AA, 193 \AA, 171 \AA, 131 \AA, and 304 \AA \ intensities in the condensation site (red box in \fig{fig6} (d)). The inverse of the AIA 304 \AA \ light curve was plotted, as cool materials resulting from the condensation manifest themselves as dark absorption features in 304 \AA \  images. As shown in \fig{fig6} (f), these light curves sequentially increase to their peaks at different times, and those with lower characteristic temperatures peak at later times, indicating the occurrence of coronal condensation due to thermal instability.
Considering the fact that the coronal-rain-related SSDs were also detected in Mg~{\sc{ii}} lines (see Section 3.2), we can conclude that the hot coronal plasma in our observations cool though $\sim$2.0 MK all the way down to $\sim$$0.01$ MK.

\subsection{Response of supersonic downflows in the sunspot umbra: localized heating of the chromosphere}
      \begin{figure*}
   \centering
   \includegraphics[width=0.8\hsize]{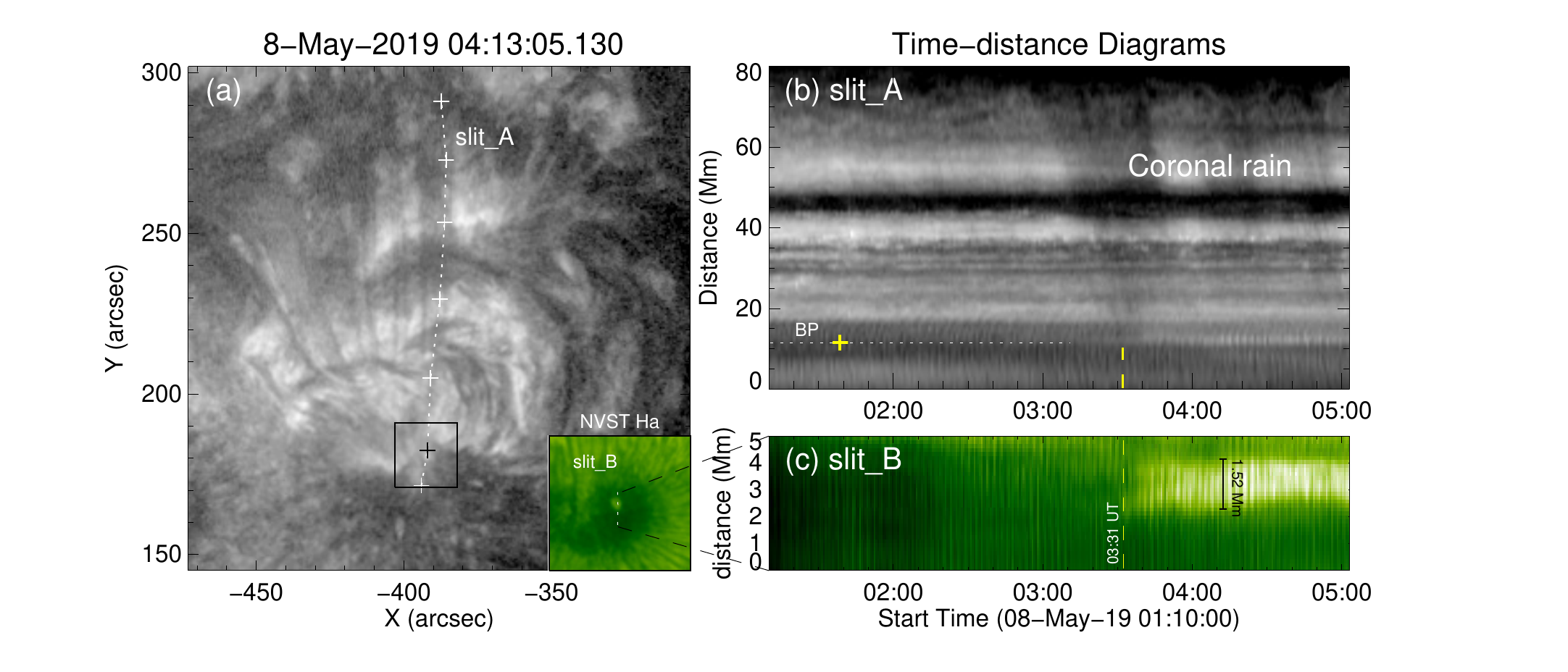}
      \caption{{Causal relationship between enhanced localized heating of the chromosphere and coronal rain flows in region B. (a) AIA 304 \AA \ image. At its right bottom corner, an NVST H$_{\alpha}$ image shows the enhanced localized heating of the chromosphere in sunspot. (b) Space-time diagram of AIA 304 \AA \ intensity for the trajectory ``Slit A" shown in panel (a). (c) Space-time diagram of NVST H$_{\alpha}$ intensity for the trajectory ``Slit B" shown in the H$_{\alpha}$ image in panel (a) . The yellow vertical dashed lines in panels (b) and (c) represent the start time (03:31 UT) of the bright dot appearance. The yellow plus signs in panel (a) and (b) denote the position of the bright dot.}
            }
         \label{fig71}
   \end{figure*}

As the downflows fall into the sunspot umbra at supersonic speeds, in H$_{\alpha}$ {line core and wing} images, we found the appearance of localized brightenings in both regions A and B during 01:00$-$06:15 UT, i.e, Rasters 2$-$4. From \fig{fig2} (a), we can see that these chromospheric brightenings only occur at two small regions (marked by the two arrows) around the footpoints of two sets of sunspot plumes. Similar brightenings can also be identified from the TR images of IRIS SJI 1400 \AA, resembling the TR bright dots or ribbons reported by \citet{2014ApJ...789L..42K}. Similar brightenings have also been reported in previous coronal rain observations \citep[e.g.,][]{2016NatSR...624319J,2021ApJ...916....5S}, which have been attributed to the heating induced by a stationary shock \citep[e.g.,][]{2015A&A...582A.116S,2016A&A...587A..20C,2021ApJ...916....5S} or a plasma collision process {\citep{2013Sci...341..251R,2015ApJ...807..142F,2017ApJ...838...15L}}.

   \begin{figure*}
   \centering
   \includegraphics[width=0.7\hsize]{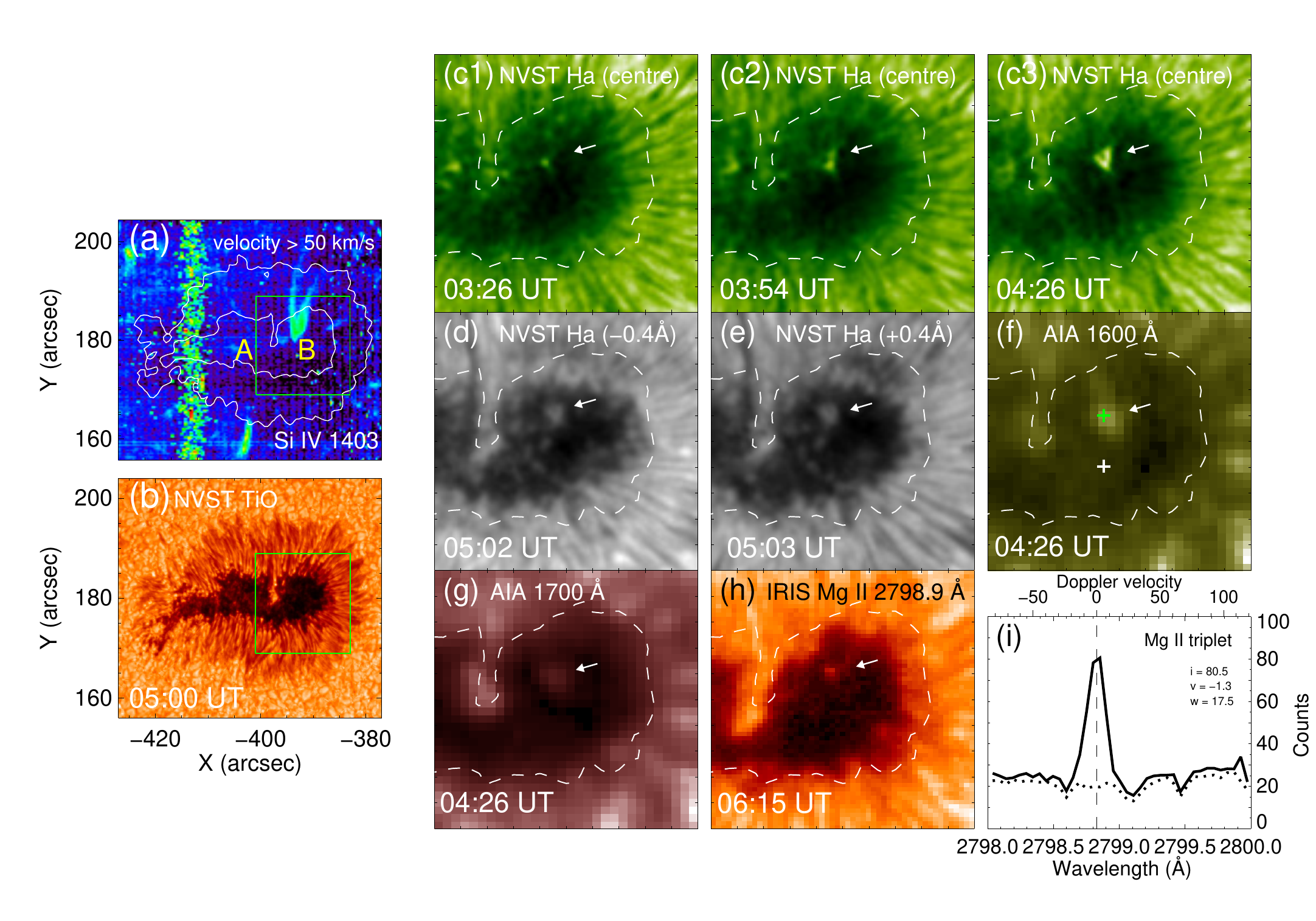}
      \caption{Enhanced localized heating of the chromosphere induced by sunspot supersonic downflows in region B. (a) IRIS Si~{\sc{iv}} downflow intensity map. (b) NVST TiO image. The green boxes indicate the FOV of the snapshots in the right part. (c) NVST H$_{\alpha}$ image sequence. (d) {NVST H$_{\alpha}$ blue wing image. (e) NVST H$_{\alpha}$ red wing image. (f)} AIA 1600 \AA \ image. (g) AIA 1700 \AA \ image. (h) Intensity map at the wavelength of 2798.90 \AA. The white dashed line marks the outer boundary of the sunspot umbra. {The white arrows mark the bright dot induced by downflows}. (i) The solid and dashed curves represent spectral profiles of Mg~{\sc{ii}} 2798.809 \AA\, averaged over nine pixels around the green and white plus signs shown in (f), respectively. {The line parameters, including the peak intensity (i), Doppler shift (v), and line width (w), are also given for the Mg II triplet.}
      }
         \label{fig7}
   \end{figure*}
{Through scrutinizing simultaneous the AIA 304 \AA \ image sequence, we found that these brightenings only appear as obvious coronal rain flows falling into the sunspot. 
\fig{fig71} shows the most prominent SSD-induced bright dot in our observations and its associated coronal rain, which was observed in region B during IRIS Raster 4. From the space-time diagrams, we can see obvious coronal rain flow towards the sunspot since 03:00 UT and impacts the latter around 03:30 UT. After the drainage of rain flow into the sunspot, the bright dot promptly appears in the umbra and then lives for more than 2 hr with a maximum diameter ($D_{dot}$) of $\sim$1.5 Mm.}

\fig{fig7} further shows the multi-wavelength response of this SSD-induced bright dot. 
From the H$_{\alpha}$  line core and wing image sequences, we can see that it first appears around 03:26 UT, and then lasts to 05:04 UT. The long-lived nature of this bright dot may imply a sustained energy deposition from the associated SSD events.
Apart from that, this bright dot is also detected in the AIA 1600 \AA \ and 1700 \AA \ images, as well as the intensity map of the Mg~{\sc{ii}} triplet line at 2798.809 \AA\, (a blend of 2798.82 \AA \ and 2798.75 \AA). This suggests that the brightening phenomenon extends to the lower chromosphere or even the temperature minimum region. From \fig{fig7} (g), we can see that the Mg~{\sc{ii}} 2798.809 \AA\, line turns into a strong emission feature at the bright dot.
Note that the Mg~{\sc{ii}} triplet lines (2791.60, 2798.82 \AA \  and 2798.75 \AA) formed in the lower chromosphere usually manifest themselves as absorption lines in most on-disk observations. They can turn into emission in case of heating in the lower chromosphere, such as in flares, UV bursts, or Ellerman bombs \citep[e.g.,][]{2014Sci...346C.315P,2015ApJ...812...11V,2015ApJ...811..139T,2016ApJ...824...96T}. 
According to the forward modelling results of \citet{2015ApJ...806...14P}, the core-to-wing ratio of the Mg~{\sc{ii}} triplet lines is well correlated to the temperature increase. In the bright dot, the core-to-wing ratio of Mg~{\sc{ii}} triplet, reaches up to 4, indicating a  temperature increase by $\sim$1500 K in the lower chromosphere \citep{2015ApJ...806...14P}. 

Assuming that SSDs fall downwards into the umbra along one flux tube and the bright dot occurs at its endpoint, we can roughly estimate the thermal and kinetic energy flux that pass through the bright dot. According to \citep{1981ApJS...45..635V}, we assume that the initial chromospheric temperature of the umbra is around 6.0$-$7.0 kK in H$_{\alpha}$ passbands. If the bright dot in the umbra undergoes a temperature increase by $\sim$1500 K, its temperature ($T_\mathrm{H_{\alpha}}$) should reach up to 7.5$-$8.5 kK. As a result, the thermal energy flux in the bright dot ($\frac{3}{2}n_ek_{\rm B}{T_\mathrm{H_{\alpha}}} \cdot v_\mathrm{SSD}$) can be estimated as 3.9$-$4.4 $\times 10^{5}$ erg cm$^{-2}$ s$^{-1}$, where $k_{\rm B}$ is Boltzmann constant and $n_{e}$ is the average density of SSDs ($\sim$ 10$^{10.4}$ cm$^{-3}$). On the other hand, the kinetic energy flux of SSDs ($\frac{1}{2}\Delta mv^2_\mathrm{SSD}$) along the flux tube can be derived as $1.9-3.4 \times 10^7$ erg cm$^{-2}$ s$^{-1}$, which is two orders of magnitude higher than the thermal energy flux in the bright dot. This suggests that a small portion of the released kinetic energy from SSDs (or falling coronal rain) is sufficient enough to cause these dot-like localized chromospheric brightenings.

\section{Discussion}
\subsection{Downflow speeds in different spectral lines}
In Section 3.2., our spectral analysis reveals that the falling coronal rain flows show almost the same speeds ($\sim$\speed{95}) in the O~{\sc{iv}} 1400/1401 \AA \ and Si~{\sc{iv}} 1403 \AA \ lines, but demonstrate a higher speed ($\sim$\speed{135}) in the Mg~{\sc{ii}} 2796 \AA \ line.
This result might be explained in terms of the multi-thermal nature of coronal rain flow or/and the geometry of magnetic structures. 
As \citet{2020PPCF...62a4016A} stated, coronal rain flows or blobs are composed of a cool, dense core and a hot, thin envelope, which have a strong co-spatial emission in chromospheric and TR lines. When a coronal rain blob falls downward towards the solar surface from the corona under the gravity, its multi-thermal components would acquire almost the same speed at a given height due to its typical small scale of 300 km width and 1 Mm length \citep{2020PPCF...62a4016A}. For the O~{\sc{iv}} and Si~{\sc{iv}} lines, they form at temperatures almost a factor of two apart, but their emission mainly come from {the enveloping TR sheath around the cool blob, as well as the wake of the rain and the compressed region ahead of the blob. In other words, {the TR emission of downflows should come from multiple atmospheric layers, effectively reducing their velocity measurement.}} Hence, downflows in O~{\sc{iv}} and Si~{\sc{iv}} lines show comparable speeds. 

{The emission of the Mg~{\sc{ii}} line mainly comes from the cool inner core of falling rain clumps, so that it gives a more accurate velocity measurement of downflows than that of TR lines. For falling rain flows, the simulations of \citet{2016ApJ...818..128O} show an increase of falling speed with higher density due to the pressure gradient restructuring. Although their simulation result is still physically unclear, the same mechanism might be operating here, leading to a higher speed in the Mg~{\sc{ii}} line.} Moreover, the gravitational acceleration may also play a role here, since the Mg~{\sc{ii}} line likely samples plasma cooled and accelerated from the Si~{\sc{iv}}- and O~{\sc{iv}}-emitting plasma at larger heights. Besides, in sunspot plumes, magnetic fieldlines are usually organized in the shape of funnels. The cross section of a funnel becomes smaller when going down from the source region of Si~{\sc{iv}} to that of Mg~{\sc{ii}}. According to Bernoulli's law, the downflow speed in Mg~{\sc{ii}} would be higher than in Si~{\sc{iv}} if its density of does not increase too much. {Using the continuity equation ($\rho v S $ = constant, where $\rho$ is the mass density, $S$ is the cross section.), we can infer that the ratio of cross section in the TR and chromospheric height is equal to 14, because the density in the chromosphere should be a factor of 10 higher than in the TR. This means that the magnetic field strength along the sunspot plumes decreases by about 14 times from the chromosphere up to the TR, due to the magnetic flux conservation.} 

\subsection{Sunspot supersonic downflows: bursty or quasi-steady}
{
It has been long hypothesized that SSDs have an association with coronal rain or prominence material, due to {their} coronal origin and chromospheric properties. However, the {existence of} long duration SSD events {(over several hours in some cases)} and their sometimes steady spectral character {question this} hypothesis. For instance,  \citet{2015A&A...582A.116S} argued that the well-known intermittency of coronal rain is inconsistent with the stable spectral nature of such long-lived steady SSDs. To date, despite that several previous observations have demonstrated that bursty SSDs result from impulsive coronal rain \citep{2014ApJ...789L..42K,2020SoPh..295...53I}, the long-lived steady SSDs have rarely be attributed to coronal rain.}
\par
{In our observations, almost all the identified SSDs appear at the footpoints of sunspot plumes and are temporally associated with episodes of coronal rain. Interestingly, the coronal-rain-induced SSDs we detected in the 5th raster of IRIS (as shown in \nfigs{fig4} (b1)-(b3)) appear as a secondary redshifted emission peak, which remains relatively stable between 18:55 UT to 19:30 UT in both TR lines and Mg~{\sc{ii}} k. Such long-lived and stable spectral character is very similar to that in \citep{2014ApJ...786..137T,2015A&A...582A.116S}. Hence, our current results provide strong evidence that steady SSDs can also be induced by falling coronal rain flows. {W}e also notice that in a coronal-rain related SSD event in \citep{2020SoPh..295...53I}, falling rain flows simultaneously result in steady SSD spectra in TR lines and bursty SSD spectra in chromospheric lines. These observations together support the view that both bursty and steady SSDs have a common origin: flows of condensation forming above sunspots due to TNE and/or thermal instability \citep{2020PPCF...62a4016A}. 
}

{It is worth noting that in \nfigs{fig4} and 6, condensed materials first accumulate as a transient prominence in a newly-formed magnetic dip region and thus form a mass reservoir available to feed a continuous and less-clumpy rain flow. Hence, the presence of magnetic dips should provide a desirable condition for the formation of steady SSDs \citep[see also][]{2021ApJ...916....5S}.
Moreover,} given that the coronal rain flows persistently drain into the sunspot along different trajectories in funnel-like magnetic structures (sunspot plumes), the funnel effect of this magnetic geometry may further reshape the less-clumpy rain at the coronal height into a more elongated and stream-like one when reaching the lower atmosphere \citep{2015ApJ...806...81A}, leading to quasi-steady redshifted spectral features. In addition, for our case, it is likely that the difference of the times when two falling blobs in the rain reach the sunspot is smaller than the sampling cadence of the IRIS slit, which is about half minute in Raster 5 (for the last slit position). This, together with the long exposure time (7.9 s), ensures that each exposure of IRIS can capture at least one falling blob. Consequently, the redshifted component is present in all exposures of IRIS. {Therefore, we suggest that in the sunspot atmosphere, the appearance of steady or bursty SSDs is not only regulated by its magnetic field topology, but {their detection} also depend{s} on the sensitivity and time resolution of IRIS raster scans. }

\subsection{Can the coronal condensation provide enough mass to feed the sunspot supersonic downflows?}
      \begin{figure}
   \sidecaption
   \includegraphics[width=\hsize]{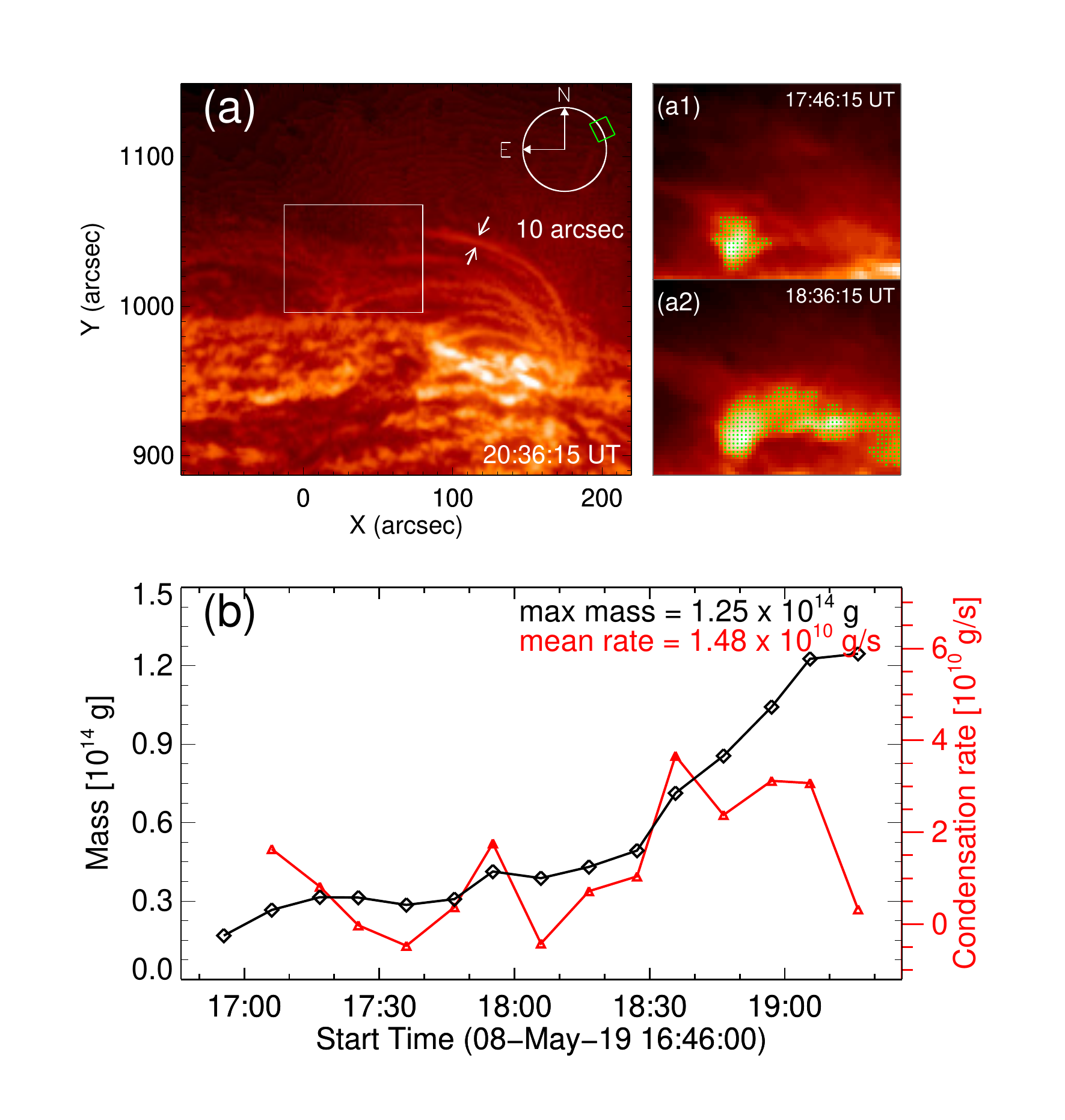}
      \caption{Source of sunspot supersonic downflows. (a) An EUVI 304 \AA \ image showing the condensation-related dip region. The solar disk and the FOV of these images are shown as the white circle and the green rectangle, respectively. The two arrows indicate the thickness for a condensation clump. (a1)-(a2) Examples of condensation identification in the white box of panel (a) at two different times. The green dots denote the pixels where condensation takes place. 
(b) Time variation of the condensation mass (black diamonds) and condensation rate (red triangles) during the period of 16:46 -- 19:26 UT (IRIS Raster 5).}
         \label{fig8}
   \end{figure}
{Apart from their steady spectral character, the mass supply for long-lived SSD events also needs to be considered.}
Focusing on an SSD event observed by IRIS, \citet{2015A&A...582A.116S} derived an electron density of 10$^{10.6\pm0.25}$ cm$^{-3}$ and a mass flux of $5 \times 10^{-7}$ g cm$^{-2}$ s$^{-1}$. Based on rough calculations, \citet{2015A&A...582A.116S} and \citet{2016A&A...587A..20C} argued that such a substantial mass flux would evacuate the overlying corona on a timescale of $\sim$100\,s to 1000\,s without supply of additional mass.

The electron density of $\sim$ 10$^{10.4}$ cm$^{-3}$ and mass flux of 5 $\times$ 10$^{-7}$ g cm$^{-2}$ s$^{-1}$ we estimated for the SSD events here are roughly consistent with previous estimations \citep{2015A&A...582A.116S,2018ApJ...859..158S,2020A&A...640A.120N}.
As mentioned above, the SSD events in our observations are clearly related to repeated quiescent coronal rain events. In particular, as shown in Section 3.2, the quasi-steady redshifted component detected during IRIS Raster 5 lasts for at least 35 minutes. Simultaneous STEREO-A/EUVI observations show that this SSD event is caused by the drainage of cool plasma originating from the coronal condensation at a high-lying (magnetic) dip region. The continuous drainage lasts for nearly 2 hrs (see \nfigs{fig4} and 5, and their associated animations), suggesting that the mass supply from coronal rain flows should be enough to sustain a long-lived SSD event. 
\par
To quantify the mass supply, we further estimated the condensation rate. 
Following \citet{2012ApJ...745L..21L}, we assumed that the EUVI 304 \AA  \ emission of coronal rain mainly comes from the prominence-corona TR (PCTR) surrounding the condensation clumps. Accordingly, using the typical electron density of a PCTR ($8 \times 10^{9}$ cm$^{-3}$), we obtained a mass density ( {1.2}$n_{e}m_{p}$) of $ {1.5} \times 10^{-14}$ cm$^{-3}$ (considering the He abundance and assuming a fully ionised plasma \citep{2020Sci...369..694Y}). Next, focusing on the dip region outlined by the white box shown in \fig{fig8} (a), we computed the apparent areas ($A_{cond}$) of condensation at different times using EUVI 304 \AA \ images taken during 16:56 $-$19:26 UT. This was achieved by identifying image pixels where the intensity is higher than 1$\sigma$ above the median intensity in the dip region. As shown in \nfigs{fig8} (a1)-(a2), this threshold ensures that we can separate and identify all brightest pixels covering condensation region from the noisy background. In addition, we assumed that there is no overlap between different condensation clumps in the LOS. 
Using the observed thickness ($\sim$ 10 arcsec) of an individual coronal rain flow (see \fig{fig8} (a)) as the possible thickness ($D_{thick}$) for the condensation clumps, the instantaneous mass of condensation at each time can be calculated as $M_{inst} =  {1.2}n_{e}m_{p}\cdot A_{cond}\cdot D_{thick}$. As shown in \fig{fig8} (b), the estimated total mass of condensation ($M_{total}$) in the dip region is $\sim$$ {1.3} \times 10^{14}$ g, and the mean condensation rate ($\dot M$) is $ {1.5} \times 10^{10}$ g s$^{-1}$. 
Assuming that the H$_{\alpha}$ bright dot in region B (see Section 3.4) represents the cross section area ($A_{loop} = \pi\cdot\frac{D_{dot}}{4}^2$) of the related flux tube, we further derived a mass transport rate ($\dot M_{tube} = A_{loop}\cdot \Delta m$) of  { $0.7 \times 10^{10}-1.2\times 10^{10}$ g s$^{-1}$}. 
Note that downflows can also be detected in the optically thick 304 \AA \ channel, thus the $F_{mass}$ of SSDs we derived from optically thin lines only gives a lower limit. So the coronal condensation can supply mass to the SSDs at most for {$2.8-4.9$} hrs ($\frac{M_{total}}{\dot M_{tube}}$), consistent with the observed durations of quasi-steady SSDs \citep[e.g.,][]{2014ApJ...786..137T,2015A&A...582A.116S}. These estimations reinforce our suggestion that coronal condensation occurring at a magnetic dip can provide enough mass to sustain a long-lived SSD event.

\subsection{How do the magnetic dips form in the high-lying magnetic structures?}
   \begin{figure*}
   \centering
   \includegraphics[width=0.7\hsize]{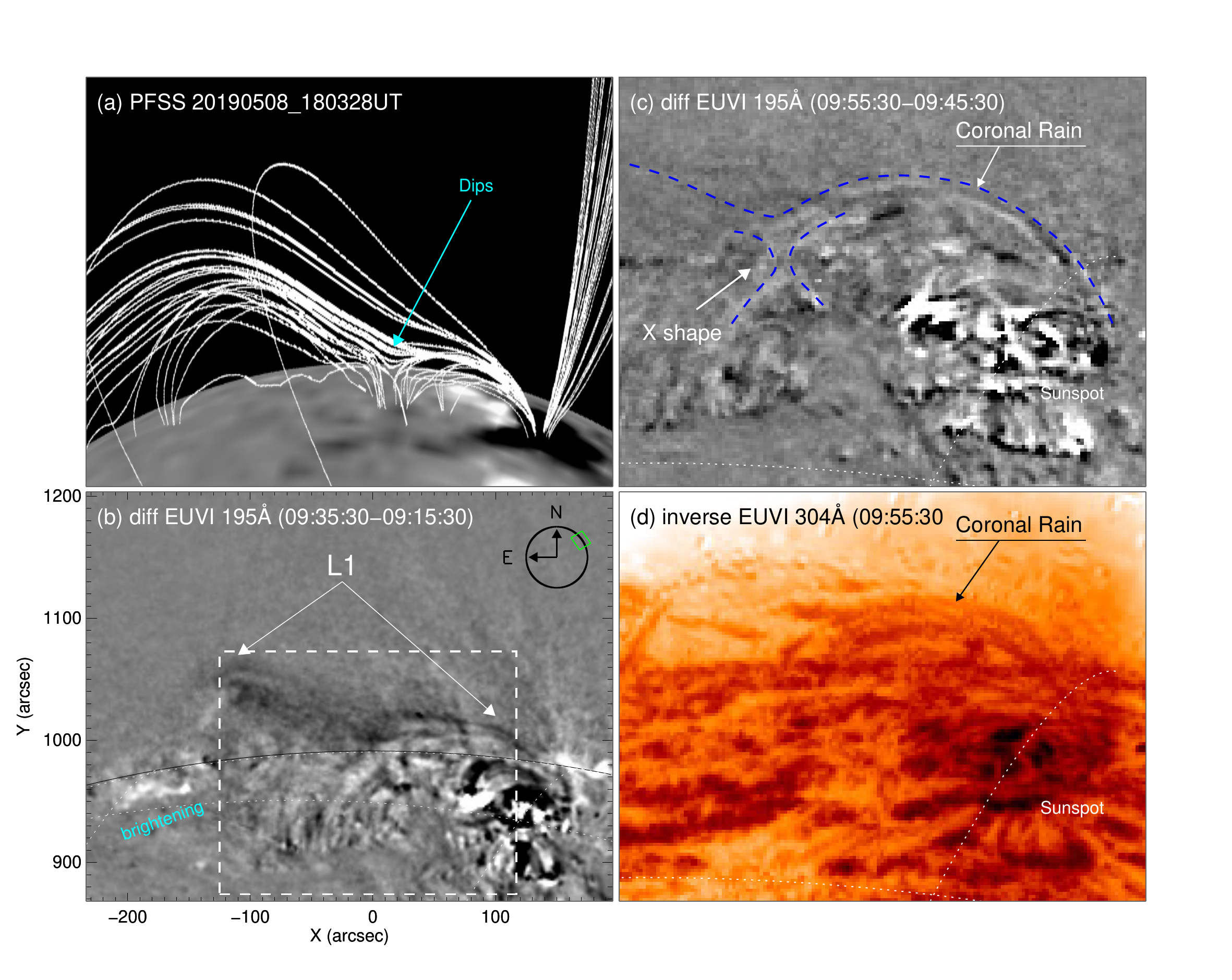} 
      \caption{Magnetic topology and related coronal rain flows near the source region of sunspot supersonic downflows. {(a) The reconstructed magnetic topology from the PFSS model. (b) The observed morphological topology in EUVI 195 \AA \ running-difference image at around 09:35 UT.} The solar disk and the FOV of these images are shown as the white circle and the green rectangle, respectively. {``L1" denotes the upward-moving coronal loops from the X-shaped structure. Near the left feet of ``L1", a weak brightening signal was observed.} (c) EUVI 195 \AA \ running-difference image and (d) 304 \AA \ image (panel (d), using reversed color table) taken at 09:56 UT. The FOV of (c) and (d) is shown as the dashed-line rectangle in (b). The coronal rain and the X-shaped structure below the dip are marked by the arrows. The dashed lines outline the possible magnetic field geometry. An animation showing the temporal evolution of panels (c) and (d) during 02:00$-$12:55 UT is available online.
      }
         \label{fig9}
   \end{figure*}
As stated in Section 3.3, the condensation of coronal plasma is apparent after the appearance of dips in the high-lying coronal structures, which is very similar to the scenario of reconnection-facilitated condensation proposed by \citet{2018ApJ...864L...4L}. 
\nfig{fig9} and its associated online animation present the topological changes of the coronal loop system in the course of condensation. In \nfig{fig9} (c) and (d), one can notice that groups of downward-concave coronal loops are continuously created near an X-shaped structure during 03:50 $-$05:30 UT. Subsequently, coronal rain appears along the upward-moving coronal loops and rapidly slides into the sunspot region, {although we should note that not all the coronal rain here is coming from the localized condensation in dips but partly from overlying (closed) loops and nearby prominences.} These EUV observations indicate that the formation of the dips is indeed closely related to the reconfiguration of coronal magnetic field lines near the X-shaped structure. Unfortunately, the low-cadence EUVI observations do not allow us to investigate the detailed reconfiguration process.

We thus reconstructed the coronal magnetic field structures using a potential field source surface (PFSS) model \citep{1969SoPh....6..442S}. 
{As shown in the \nfig{fig9} (a) and (b), the reconstructed X-shaped structure roughly agrees with the global topological morphology observed in the EUVI 195 \AA \ running-difference image at around 09:35 UT, in which obvious magnetic dips exist along the downward-concave coronal loops above the X-shaped structure. 
Note that near the left leg of the downward-concave coronal loops, a weak brightening signal appears at around 09:35 UT. This indicates possible plasma heating and evaporation in the reconfiguration of coronal magnetic field lines, which are thought to be a key condition for the formation of coronal rain and prominences in numerical simulations \citep[e.g.,][]{2014ApJ...792L..38X,2015ApJ...807..142F,2016ApJ...823...22X,2021ApJ...913L...8H}.}
\par
   \begin{figure*}
   \centering
   \includegraphics[width=.65\hsize]{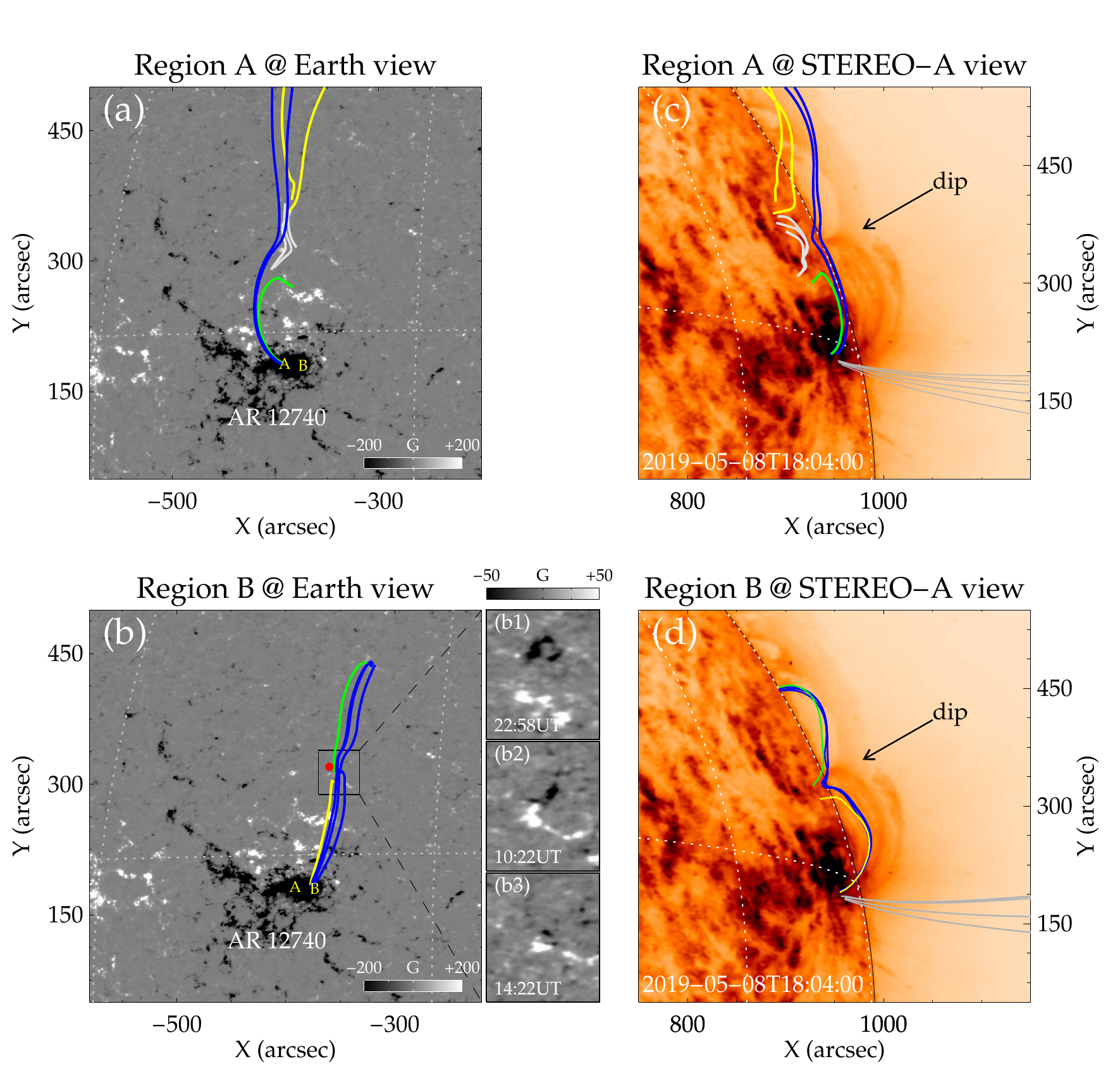}
      \caption{Magnetic topology. (a) and (b) Extrapolated magnetic field lines related to the sunspot supersonic downflows in regions A and B projected onto SDO/HMI light-of-sight (LOS) magnetograms. (c) and (d) are similar to (a) and (b), but projected onto STEREO/EUVI 304 \AA \ images with reversed colors. Three inset panels (b1)-(b3) show a zoomed-in view of the magnetic field in the black box in (b) at three different times. The red dot in (b) marks the location of the magnetic dip, which is also  shown in \fig{fig6}.
              }
         \label{fig10}
   \end{figure*}

\nfig{fig10} further shows several extrapolated field lines projected onto SDO/HMI and STEREO-A/EUVI images. We can see obvious magnetic dips on the extrapolated high-lying field lines (the blue lines) {at two different spatial regions where coronal condensation take place (as shown in \fig{fig4} and \fig{fig6}).}  
For region A, the related magnetic field configuration is very similar to the magnetic setting in the reconnection-facilitated condensation scenario proposed by \citep{2018ApJ...864L...4L}. The higher-lying curved structures (blue lines) may move downward and reconnect with the lower-lying loops (white lines). As a result, two sets of new magnetic structures (green and yellow lines) form at opposite sides of the lower-lying loops. The regions separating these different magnetic structures appear as an X-shaped feature. As suggested in \citet{2020ApJ...905...26L}, due to the downward motion of the higher-lying structures, a magnetic dip would form above the X-shaped structure. 
For region B, the related magnetic field configuration is more like the common magnetic setting of solar prominence/filament formation in previous numerical simulations \citep[e.g.,][]{2014ApJ...792L..38X,2017ApJ...845...12K} and observations \citep[e.g.,][]{1989ApJ...343..971V,2016ApJ...816...41Y,2021ApJ...921L..33Y,2018ApJ...869...78C,2021ApJ...919L..21L}. We found two sets of adjacent magnetic arches (yellow and green lines) rooted at two opposite-polarity photospheric flux concentrations (see \nfigs{fig10} (b1)$-$(b3)). The two flux concentrations are found to converge and cancel with each other during 09:00 UT$-$15:00 UT, likely indicating magnetic reconnection. This process results in the formation of (sheared) overlying loops (blue lines) with a magnetic dip.

\subsection{Interpretation}
   \begin{figure}
   \centering
   \includegraphics[width=0.65\hsize]{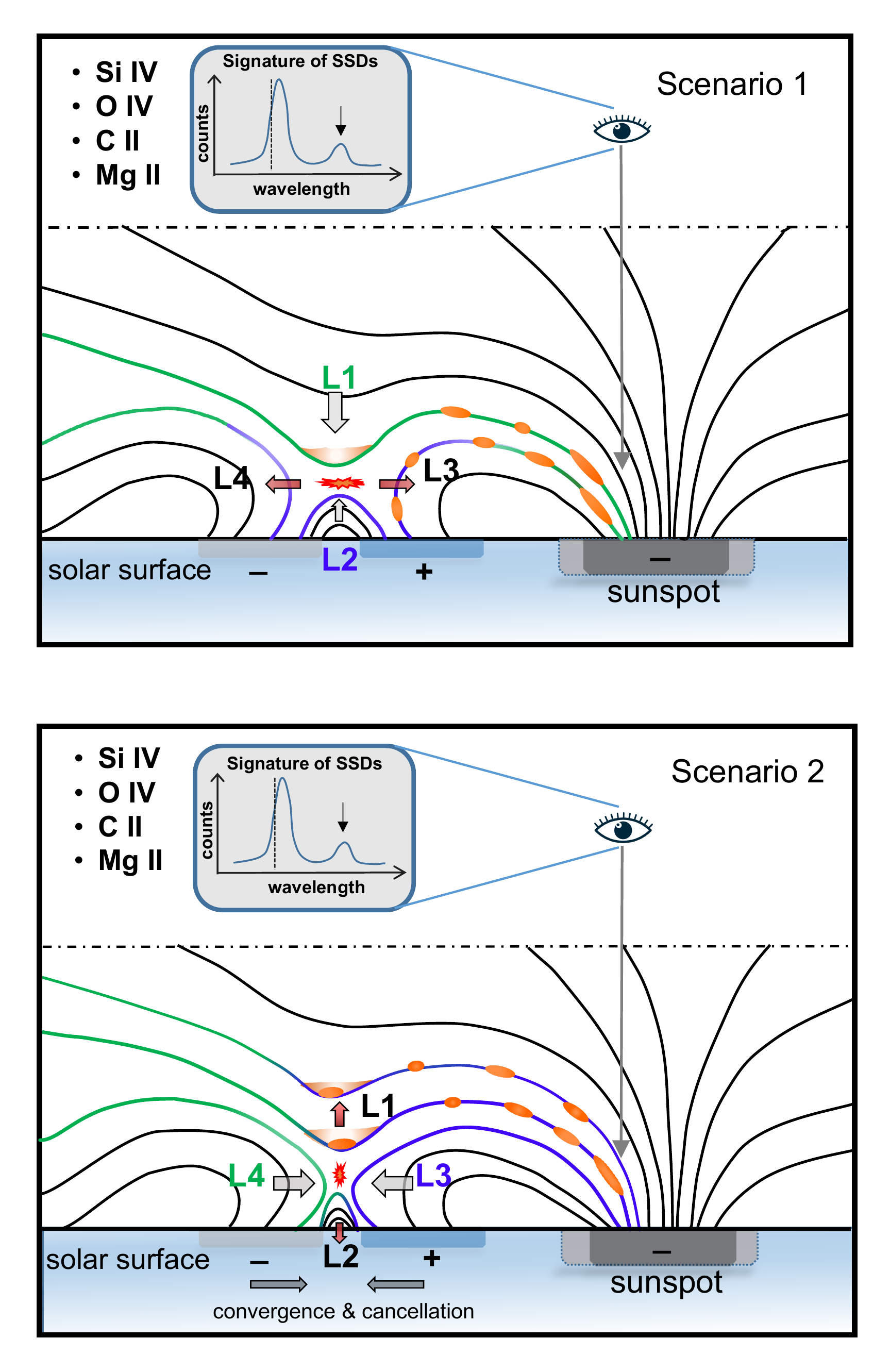}
      \caption{Two possible scenarios for the magnetic dip formation, subsequent coronal condensation, and supersonic downflows. The ``+" and ``-" signs denote the positive- and negative- polarity magnetic fields on the solar surface, and the solid lines represent magnetic field lines. Near the reconnection site (red explosive feature), four groups of field lines, ``L1", ``L2", ``L3" and ``L4", are highlighted by green and blue solid lines. The grey arrows indicate the converging motion of existing loops to the reconnection site, {while the red arrows mark the diverging motion of newly formed loops from the reconnection site.} The brown shadowed regions represent coronal rain clumps. Note that ``L4" in the top panel and ``L2" in the bottom panel may also exhibit coronal rain.
                   }                   
         \label{fig11}
   \end{figure} 
   
In \fig{fig11}, two schematic diagrams are given to explain {the magnetic dip formation, subsequent coronal condensation, and quasi-steady supersonic downflows.} The top one supports the reconnection-facilitated condensation scenario proposed by \citet{2018ApJ...864L...4L}. In this scenario, magnetic reconnection occurs between the L1 and L2 field lines, resulting in two sets of newly reconnected loops (L3 and L4) retracting from the reconnection site. Due to the downward motion, the higher-lying magnetic structure L1 will curve towards the magnetic null point and a magnetic dip is formed. As more field lines in the L1 structure move downward and reconnect with L2, a low pressure region would form around the dip region after episodes of reconnection. Then plasma will accumulate in the dip region. However, the bottom one is quite similar to the formation process of filaments/prominences. In this scenario, two sets of loops, L3 and L4, move together, and their reconnection leads to high-lying magnetic structures L1 and small lower-lying loops L2. The newly formed L1 structure naturally has a magnetic dip, providing a favored site for plasma accumulation. 

As plasma gradually accumulates in the dip, the cooling and condensation of hot plasma would inevitably commence most likely due to thermal instability. In this course, intensities of EUV passbands sampling cooler plasma would peak at later times. Subsequently, cool materials with temperatures of $\sim$0.01-0.1 MK would form and fill up the dip, and then slide down towards the sunspot region along the L1 (and also L3) field lines due to the gravity. As a result, from the STEREO-A viewing angle, the drainage of cool materials would be observed as off-limb quiescent coronal rain activity. While from the Earth viewing angle, the falling cool plasma would be detected as spectral signatures of SSDs near the footpoints of sunspot plumes by IRIS.

Even though the magnetic dip may originate from two different reconnection processes, our observations strongly support that reconnection facilitates the formation of magnetic dips. {These two scenarios may differentiate each other mainly by the motion of the high-lying field lines. In the former scenario, there should be field lines {moving} down towards the X-point, while in the latter the field lines should {move} up {(as inferred from the apparent motion of coronal structures)}. Both the upward and downward motion of the high-lying field lines can be seen in the animations of \nfig{fig5} and \nfig{fig9}, respectively.} 
Moreover, in the first scenario of \fig{fig11}, cool materials resulting from the condensation can also slide towards the sunspot region along the newly-formed loops L3. A similar feature was indeed observed in our case (see \nfigs{fig5} and 8). Therefore, {any of this two scenarios} cannot be ruled out with current observations.
\section{Conclusions}
We have conducted a case study on the origin of sunspot supersonic downflows (SSDs) based on joint spectroscopic and imaging observations with multiple instruments. 
{We found that almost all the identified SSDs in IRIS raster scans appear at the footpoints of sunspot plumes and are temporally associated with episodes of dot-like chromospheric heating events inside the sunspot umbra. In IRIS spectra, these fast downflows can be simultaneously detected as secondary emission peaks in several transition region and chromospheric lines. The average density, Doppler velocity and mass flux of these SSDs are of the order of 10$^{10.4} $ cm$^{-3} $, \speed{100} and $ 5 \times 10^{-7}$ g cm$^{-2}$ s$^{-1}$, respectively. Our dual-perspective EUV imaging observations reveal that a large-scale closed magnetic loop system spans from remote regions to the sunspot region, in which repeated coronal rain form at different regions and flow towards the sunspot, thus causing these SSDs.} 

{In particular, a well-observed coronal condensation in the 5th raster of IRIS occurs in a magnetic dip region that newly forms as a result of magnetic reconnection. With the presence of this dip region, cool material that condensed from the hot corona accumulates as a transient prominence, which thus serves as a desirable mass provider and results in a long-lived steady SSD event via a long-lasting continuous rain flow.}  
The SSD-related drainage of coronal rain and the coronal rain-induced chromospheric heating both last for more than 2 hrs, suggesting a substantial mass supply by the coronal condensation. The estimated total mass of condensation ($ {1.3} \times 10^{14}$ g) and condensation rate ($ {1.5} \times 10^{10}$ g s$^{-1}$) have been found to be large enough to sustain a quasi-steady SSD event, which has a mass transport rate of  {$0.7 \times 10^{10}-1.2\times 10^{10}$ g s$^{-1}$}. 
Based on imaging observations and a magnetic field extrapolation, two possible reconnection scenarios have been proposed to explain the magnetic dip formation and subsequent coronal condensation. Similar to the recent finding of \citet{2018ApJ...864L...4L} and \citet{2019ApJ...874L..33M}, our observations support that apart from the thermal evolution in the coronal magnetic structures, the magnetic topological evolution also plays a vital role in the onset of coronal condensation.

\par
Our observations cannot exclude the possibility that siphon flows may also feed SSDs. Future investigations should be conducted to examine this possibility. Statistical studies should also be performed to investigate how common this reconnection-facilitated condensation is for the origin of SSDs.

\begin{acknowledgements}
IRIS is a NASA Small Explorer mission developed and operated by LMSAL with mission operations executed at NASA Ames Research center and major contributions to downlink communications funded by ESA and the Norwegian Space Center. AIA and HMI are instruments onboard the Solar Dynamics Observatory, a mission for NASA's Living With a Star program. STEREO is a NASA mission. The H$_{\alpha}$  {and TiO} data used in this paper were obtained with the New Vacuum Solar Telescope in Fuxian Solar Observatory of Yunnan Astronomical Observatories, CAS. This work was supported by NSFC grants 11825301, 12073042 and 11790304. H.C.C. acknowledges supports by the National Postdoctoral Program for Innovative Talents (BX20200013) and the China Postdoctoral Science Foundation (2020M680201). Z. Y. H. acknowledges the support by the China Postdoctoral Science Foundation (2021M700246). This project has received funding from the European Research Council (ERC) under the European Union’s Horizon 2020 research and innovation programme (grant agreement No 695075). We thank the referee, \textbf{Dr. P. Antolin}, for his very helpful comments and constructive suggestions, and H. C. C. thanks Dr. E. Landi for helpful discussion.
\end{acknowledgements}


%
%

\end{document}